\documentclass[modern]{aastex701}

\usepackage{amsmath}

\newcommand{\kms}{${\rm km~s}^{-1}$}
\newcommand{\kmsmpc}{${\rm km~s}^{-1}~\rm{Mpc}^{-1}$}

\newcommand{\dm}{$\Delta{\rm m}_{15}(B)$}
\newcommand{\sbv}{$s_{BV}$}

\shorttitle{Fast-Declining Type Ia Supernovae}
\shortauthors{Phillips et al.}

\begin{document}

\title{Carnegie Supernova Project: Fast-Declining Type Ia Supernovae as Cosmological Distance Indicators\footnote{This paper includes data 
gathered with the 6.5 meter Magellan telescopes at Las Campanas Observatory, Chile.}}

\author[0000-0003-2734-0796]{M.~M.~Phillips}
\affiliation{Carnegie Observatories, Las Campanas Observatory, Casilla 601, La Serena, Chile}
\email[show]{mmp@carnegiescience.edu} 

\author[0000-0002-9413-4186]{Syed~A.~Uddin}
\affil{American Public University System, 111 W. Congress St., Charles Town, WV 25414, USA}
\affil{Center for Astronomy, Space Science and Astrophysics, Independent University, Bangladesh, Bashundhara R/A, Dhaka 1245, Bangladesh}
\email{saushuvo@gmail.com} 

\author[0000-0003-4625-6629]{Christopher R.~Burns}
\affiliation{Observatories of the Carnegie Institution for Science, 813 Santa Barbara St., Pasadena, CA 91101, USA}
\email{cburns@carnegiescience.edu}

\author[0000-0002-8102-181X]{Nicholas~B.~Suntzeff}
\affiliation{George P. and Cynthia Woods Mitchell Institute for Fundamental Physics and Astronomy, Texas A\&M University,
Department of Physics and Astronomy,  College Station, TX 77843, USA}
\email{nsuntzeff@tamu.edu}

\author[0000-0002-5221-7557]{C.~Ashall}
\affiliation{Institute for Astronomy, University of Hawai’i at Manoa, 2680 Woodlawn Dr., Hawai’i, HI 96822, USA}
\email{cashall@hawaii.edu}

\author[0000-0001-5393-1608]{E. Baron}
\affiliation{Planetary Science Institute, 1700 East Fort
  Lowell Road, Suite 106,Tucson, AZ 85719-2395, USA}
\affiliation{Hamburger Sternwarte, Gojenbergsweg 112, 21029 Hamburg, Germany}
\affiliation{Homer L.~Dodge Department of Physics and
  Astronomy, University of Oklahoma, 440 W. Brooks, Rm 100, Norman, OK
  73019-2061, USA}
\email{ebaron@psi.edu}

\author[0000-0002-1296-6887]{L.~Galbany}
\affiliation{Institute of Space Sciences (ICE, CSIC), Campus UAB, Carrer de Can Magrans, s/n, E-08193 Barcelona, Spain.}
\affiliation{Institut d’Estudis Espacials de Catalunya (IEEC), E-08034 Barcelona, Spain.}
\email{l.g@csic.es}

\author[0000-0002-4338-6586]{P.~Hoeflich}
\affiliation{Department of Physics, Florida State University, 77 Chieftan Way, Tallahassee, FL  32306, USA}
\email{phoeflich77@gmail.com}

\author[0000-0003-1039-2928]{E.~Y.~Hsiao}
\affiliation{Department of Physics, Florida State University, 77 Chieftan Way, Tallahassee, FL  32306, USA}
\email{yichi.hsiao@gmail.com}

\author[0000-0003-2535-3091]{Nidia~Morrell}
\affiliation{Carnegie Observatories, Las Campanas Observatory, Casilla 601, La Serena, Chile}
\email{nmorrell@carnegiescience.edu}

\author[0000-0003-0554-7083]{S.~E.~Persson}
\affiliation{Observatories of the Carnegie Institution for Science, 813 Santa Barbara St., Pasadena, CA 91101, USA}
\email{persson@carnegiescience.edu}

\author[0000-0002-5571-1833]{Maximilian Stritzinger}
\affiliation{Department of Physics and Astronomy, Aarhus University, Ny Munkegade 120, DK-8000 Aarhus C, Denmark}
\email{max@phys.au.dk}

\author[0000-0001-6293-9062]{Carlos~Contreras}
\affiliation{Carnegie Observatories, Las Campanas Observatory, Casilla 601, La Serena, Chile}
\email{ccontreras@carnegiescience.edu}

\author[0000-0003-3431-9135]{Wendy~L.~Freedman}
\affiliation{Department of Astronomy and Astrophysics, University of Chicago, 5640 S. Ellis Ave, Chicago, IL 60637, USA}
\email{wfreedman@uchicago.edu}

\author[0000-0002-6650-694X]{Kevin~Krisciunas}
\affiliation{George P. and Cynthia Woods Mitchell Institute for Fundamental Physics and Astronomy, Texas A\&M University,
Department of Physics and Astronomy,  College Station, TX 77843, USA}
\email{krisciunas@physics.tamu.edu}

\author[0000-0001-8367-7591]{S. Kumar}
\affiliation{Department of Astronomy, University of Virginia, 530 McCormick Rd, Charlottesville, VA 22904, USA}
\email{sahanak@gmail.com}

\author[0000-0002-3900-1452]{J. Lu}
\affiliation{Department of Physics and Astronomy, Michigan State University, East Lansing, MI 48824, USA}
\email{lujingeve158@gmail.com}

\author[0000-0001-6806-0673]{Anthony L. Piro}
\affiliation{Observatories of the Carnegie Institution for Science, 813 Santa Barbara St., Pasadena, CA 91101, USA}
\email{piro@carnegiescience.edu}

\author[0000-0002-9301-5302]{M. Shahbandeh}
\affiliation{Space Telescope Science Institute, 3700 San Martin Drive, Baltimore, MD 21218, USA}
\email{mshahbandeh@stsci.edu}

\begin{abstract}

In this paper, the suitability of fast-declining Type~Ia supernovae (SNe~Ia) as cosmological standard candles is examined utilizing a Hubble
Flow sample of 43 of these objects observed by the Carnegie Supernova Project (CSP).  We confirm previous suggestions that fast-declining SNe~Ia 
offer a viable method for estimating distances to early-type galaxies when the color-stretch parameter, \sbv, is used as a measure of the light 
curve shape.  As a test, we employ the Tripp method, which models the absolute magnitude at maximum as a function of light curve shape 
and color.  We calibrate the sample using 12 distance moduli based on published Infrared Surface Brightness Fluctuations to derive a 
value of the Hubble constant that is in close agreement with the value
obtained for the full sample of CSP SNe~Ia using the same methodology.
We also develop a new and simple method of estimating the distances of fast decliners based only on their colors 
at maximum (and not light curve shape) and find that it leads to similar results as with using the Tripp method.
\textit{This ``Color'' technique is a 
powerful tool that is unique to fast-declining SNe~Ia.}  We show that the colors of the fast decliners at maximum light are strongly
affected by photospheric temperature differences and not solely due to dust extinction, and provide a physical rationale for this effect.

\end{abstract}


\section{Introduction}
\label{sec:intro}

Type~Ia supernovae (SNe~Ia) have been at the core of measurements of the local rate of expansion of the universe ever since the Cal\'{a}n/Tololo
project demonstrated that these objects are superb high-precision cosmological standard candles \citep{hamuy96a,hamuy96b}. Relying heavily on
the Hubble flow defined by SNe~Ia, \citet{freedman01} measured a Hubble constant, $H_0$, of $72 \pm 8 $~\kmsmpc\ based on Cepheid distances to
host galaxies measured with the Hubble Space Telescope (HST).  Since then, nearly all direct measurements of $H_0$ using SNe~Ia
have yielded values consistent with a range of $\sim$70--75~\kmsmpc\ \citep[e.g, see][]{garnavich23,riess24,uddin24,freedman25}.  In contrast, 
the value of $H_0$ inferred from observations of the Cosmic Microwave Background (CMB) invoking the concordance Lambda-Cold Dark Matter 
($\Lambda$CDM) cosmological model \citep{planck16}, or from baryon acoustical oscillations calibrated by the CMB \citep{guo25}, is in the range
of 67--68~\kmsmpc.  Using HST observations of Cepheid variables to calibrate SNe~Ia, 
this difference between the current expansion rate and that predicted from the
early universe, commonly referred to as the 
``Hubble tension,'' was found by \citep{riess22} to be at the $5\sigma$ level, a conclusion also reached by a more recent community-based review of all available data sets \citep{2025arXiv251023823H}.
Among possible explanations for the tension are new physics beyond the $\Lambda$CDM model, cosmic ``supervoids,'' modified
gravity, or problems with the calibration of the local standard candles used to measure $H_0$ or the CMB data \citep[see][for an extensive 
review of the Hubble tension problem]{hu23}.
Note, however,
that \citet{freedman25}, did not confirm such a large discrepancy using a combined HST and James Web Space Telescope sample of 
Tip of the Red Giant Branch (TRGB) stars as calibrators, and urged caution before 
abandoning the $\Lambda$CDM model.

Fast-declining SNe~Ia, which are linked with the \citet{branch06} ``cool'' (CL) spectroscopic class, occur predominantly in massive galaxies 
with old stellar populations \citep[see, e.g.,][]{hamuy00,neill09,ashall16a,rigault20} and have rates in the local universe that are among the
lowest of all SNe~Ia \citep[see, e.g.,][]{neill09,sharon22}. The prototype of this subclass, SN~1991bg, was extensively observed by 
\citet{filippenko92b} and \citet{leibundgut93}.  At $B$-band maximum, 1991bg-like SNe are as much as three magnitudes fainter than a normal 
SN~Ia\footnote{In this paper, a ``normal SN~Ia'' refers to the ``core-normal'' definition of \citet{branch06}.}.  This property, along with the
widely-held opinion that 1991bg-like SNe are ``peculiar,'' has made fast-declining SNe~Ia relatively unattractive for cosmological studies.
Moreover, both the \dm\ parameter \citep{phillips93} and the {\it x1} stretch parameter of the widely-used \texttt{SALT2} light curve fitter 
\citep{guy07}, break down as reliable discriminators of light curve morphologies and absolute magnitudes for SNe~Ia with \dm~$\gtrsim 1.7$~mag 
or {\it x1}~$\lesssim -1.7$ \citep{phillips12,burns14}.

\begin{figure*}[t]
\epsscale{.6}
\plotone{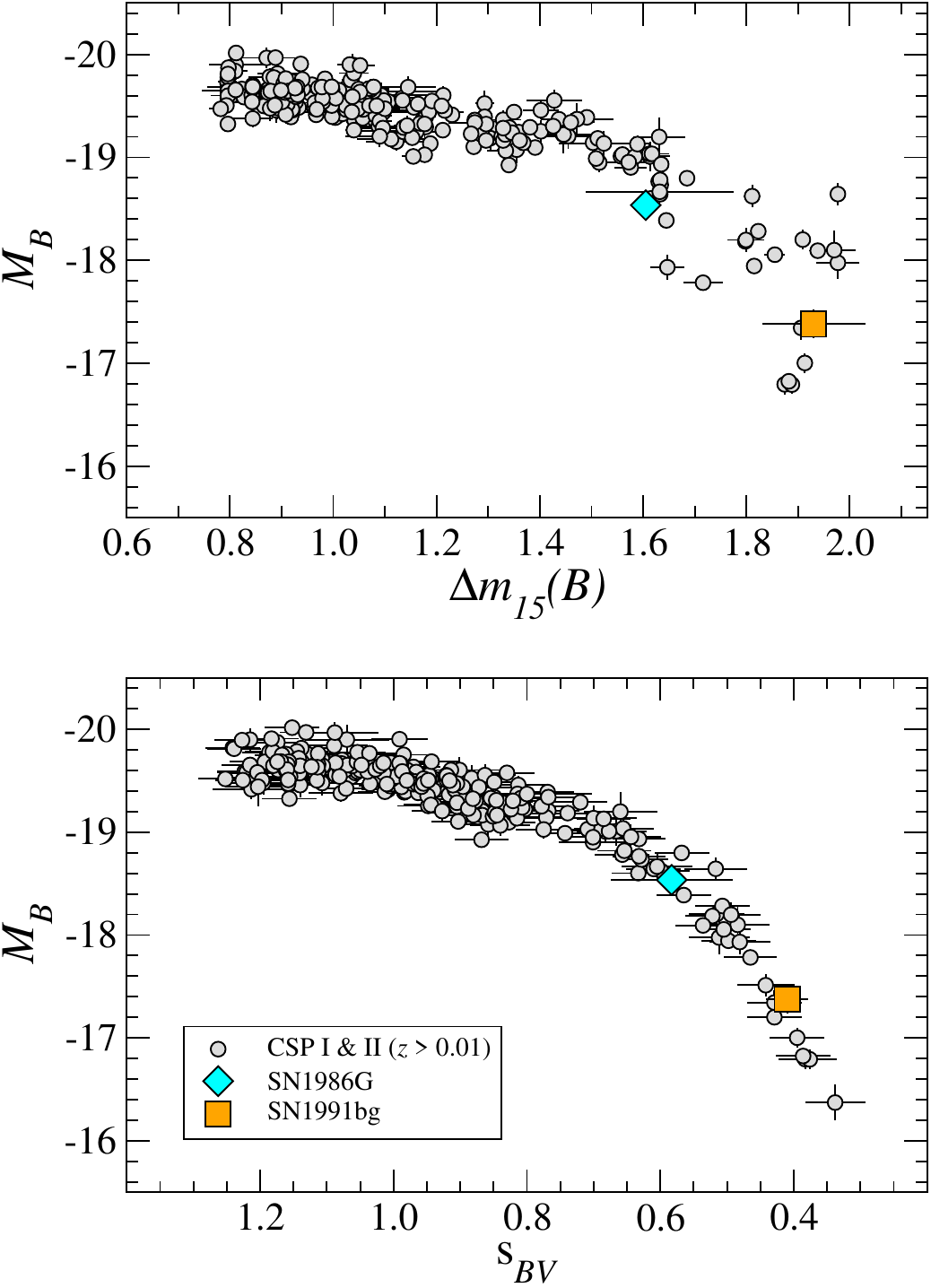} 
\caption{(above) Absolute magnitudes, $M_B$, plotted versus the light curve decline rate parameter, \dm, of SNe~Ia observed by the CSP~I and 
CSP-II with $z_{\rm {CMB}} > 0.01$.  The K-corrected magnitudes have been adjusted for Milky Way dust reddening, and also for host galaxy dust
extinction using the intrinsic color analysis described in detail by \cite{burns18}.  Distance moduli were derived from $z_{\rm {CMB}}$ 
assuming standard $\Lambda$CDM cosmology and a fixed Hubble constant $H_0 = 72$~\kmsmpc, density parameter $\Omega_m = 0.27$, and cosmological 
constant parameter $\Omega_\Lambda = 0.73$.  Plotted for reference are points for SN~1991bg and the ``transitional'' SN~1986G.  For SN~1991bg,
we used the SBF distance modulus for the host, NGC~4374, from \citet{tonry01}, subtracting 0.16~mag \citep{jenson03} to put this on the 
\citet{freedman01} scale.  For SN~1986G, we adopted the revised SBF distance of \citet{ferrarese07}. (below) Same data plotted instead versus
the color-stretch parameter, \sbv.}
\label{fig:MB_v_sBV}
\end{figure*}

Crucially, \citet{burns14} found that the latter problem was solved by replacing \dm\ or $x1$ by a new, dimensionless ``color-stretch'' 
parameter, \sbv, defined as the time difference between $B$-band maximum and the reddest point of the ($B - V$) color divided by 30 days.
Figure~\ref{fig:MB_v_sBV} shows absolute $B$ magnitudes of SNe~Ia observed by the CSP~I \citep{hamuy06} and CSP~II \citep{phillips19} plotted 
against \dm\ (above) and \sbv (below). Also plotted are SN~1991bg and the ``transitional'' SN~1986G.\footnote{Transitional SNe~Ia are defined 
by \citet{hsiao15} as lying on the ``fast-declining edge'' of the luminosity-decline rate relation for normal SNe~Ia.} As is evident in the
upper plot, the points form an ordered sequence until \dm~$\sim1.6$~mag, at which point the scatter in absolute magnitude increases 
significantly.  However, when plotted against \sbv\ in the lower plot, the sequence of luminous SNe~Ia transitions smoothly into a ``tail'' of
events dropping by nearly three magnitudes and extending to $s_{BV} \sim 0.35$. In this diagram, the fast-declining SNe~Ia no longer appear to
be ``peculiar.''

Recently, \citet{hoogendam22} and \citet{graur24} have argued that fast-declining SNe~Ia may provide a valuable independent measurement of
$H_0$ when \sbv\ is used to parameterize the light curve shape.  However, since these SNe occur preferentially in early-type galaxies, which 
are mostly devoid of classical Cepheid variables, calibration of $H_0$ using fast-declining SNe~Ia will require the use of distance indicators 
such as Infrared Surface Brightness Fluctuations \citep[IR~SBF;][]{jensen21,garnavich23}, the Tip of the Red Giant Branch 
\citep[TRGB;][]{freedman21}, and/or the Planetary Nebula Luminosity Function \citep[PNLF;][]{feldmeier07} that are compatible with old stellar
populations.  Rather than this being a weakness, \citet{graur24} pointed out that an independent measurement of $H_0$ via these non-Cepheid 
methods would be a valuable alternative test of the current Hubble tension controversy.

The primary objective of this paper is to explore the luminosity-width and color--width relations for fast-declining SNe~Ia using optical and 
near-infrared photometry obtained by the CSP.  We confirm that this subset of SNe~Ia is well-behaved when \sbv\ is used as the stretch 
parameter, and also find that the color-width relation alone is an equally effective tool for measuring cosmological distances to subluminous 
events.  As a test, we derive the Hubble constant using exclusively the set of CSP fast-declining Hubble flow SNe~Ia calibrated via the IR~SBF 
technique.  In \S\ref{sec:sample}, details of the SN samples employed in this study are given along with maximum-light magnitudes and \sbv\ 
values derived from fits to the light curves. Next, in \S\ref{sec:analysis}, the dependence of absolute magnitude on \sbv\ and color is 
explored.  In \S\ref{sec:hubble}, the widely-used \citet{tripp98} method and a novel ``Color'' technique are employed to measure the Hubble 
constant, and in \S\ref{sec:tripp_parameters} we explore in detail the interpretion of the Tripp method color parameter, $\beta$.  Finally, in 
\S\ref{sec:conclusions}, our conclusions are briefly discussed.

\section{Samples}
\label{sec:sample}

\subsection{CSP}
\label{sec:sample_CSP}

\begin{figure*}[t]
\epsscale{.8}
\plotone{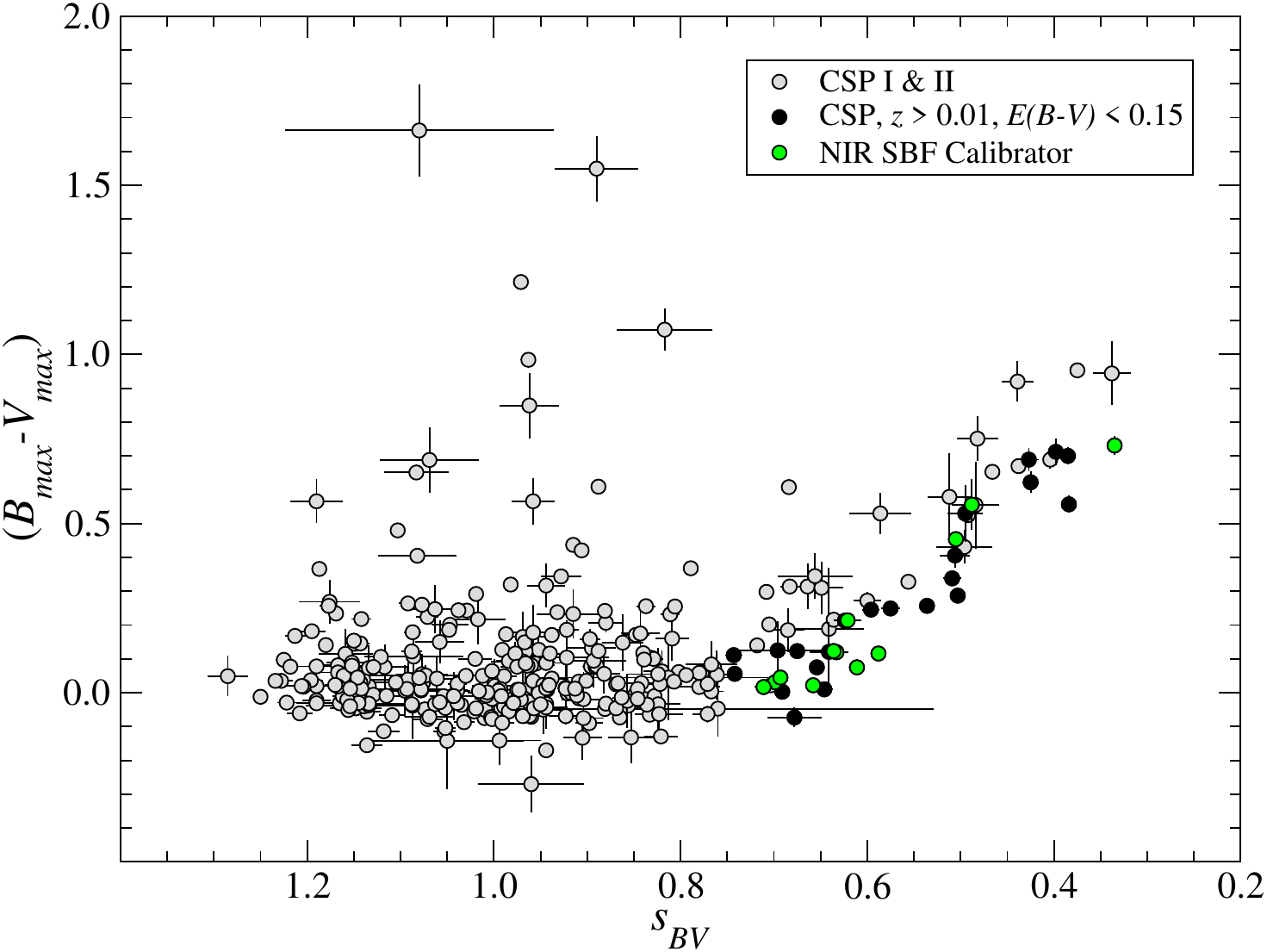} 
\caption{Observed ($B_{max} - V_{max}$) pseudocolors for CSP-I and CSP-II SNe~Ia plotted versus \sbv. The black points correspond to the
fast-declining, \sbv~$< 0.75$, subsample of CSP SNe with $z_{\rm {CMB}} > 0.01$ and with host galaxy dust reddenings $E(B-V) < 0.15$~mag as
estimated via an intrinsic color analysis like that described by \cite{burns18}.  Also plotted as green points are the IR~SBF calibrator 
SNe~Ia. Note that the colors for all objects in this plot are corrected for Milky Way dust reddening, but with no correction for host galaxy
dust reddening.}
\label{fig:BV_v_sBV}
\end{figure*}

To study the luminosity-width and color--width relations, we employ a homogeneous sample of fast-declining SNe~Ia observed by the CSP
\citep{hamuy06,phillips19} between 2004--2014 for which precise photometry was obtained
with well-characterized telescope/filter/detector
systems at the Las Campanas Observatory in Chile.  
Optical and near-infrared photometry in $uBgVriYJH$ filters for the CSP-I sample
of 123 SNe~Ia was published by \citet{contreras10}, \citet{stritzinger11}, and \citet{krisciunas17,krisciunas17_erratum}, and a paper 
presenting light curves for an additional 214 SNe~Ia comprising the CSP~II sample is currently in preparation.
We choose to define ``fast-declining'' 
SNe~Ia as those having \sbv~$< 0.75$, which includes both 1991bg-like events and transitional SNe~Ia \citep{hsiao15}.  This definition 
corresponds approximately to the point in the luminosity-width relation where the curvature begins to significantly steepen to fainter 
magnitudes (see Figure~\ref{fig:MB_v_sBV}), and is also where the observed ($B_{max} - V_{max}$) pseudocolors\footnote{Note that 
($B_{max} - V_{max}$) is a ``pseudocolor'' since it does not represent the actual color of the SN at any specific time, but is the 
difference between the $B$ and $V$ magnitudes at maximum light, which occur at slightly different epochs.} begin a nearly linear
increase (see Figure~\ref{fig:BV_v_sBV}). Limiting the sample to \sbv~$< 0.75$ includes ~80\% of the CSP SNe classified as Branch CL by
\citet{folatelli13} and \citet{morrell24}, and excludes all but four SNe classified as Branch CN or BL.
Decreasing the cutoff to \sbv~$< 0.70$ would eliminate all but one of the latter objects, 
but would also exclude five CL SNe.  Increasing the cutoff to
\sbv~$< 0.80$ would add three CL SNe, but would increase the number of CN and BL
types to eight.  Hence, a cutoff of \sbv~$< 0.75$ provides an optimal sample of Branch CL SNe.  
Note that this cutoff does not take into account errors in the \sbv\
values since these are generally small and would move only four SNe in the full CSP sample
either above or below the \sbv~$< 0.75$ cutoff.

A total of 27 SNe~Ia from the CSP-I and 28 SNe~Ia from the CSP-II meet the above definition.  Photometry for one of the CSP-II SNe~Ia, 
iPTF13dym, appears to have been compromised by poor host-galaxy subtraction due to the location of the SN near the nucleus of its host. 
Hence, we exclude this object from this analysis, giving a final total of 54 CSP SNe~Ia with \sbv~$< 0.75$.
The multi-filter light curves 
for each object were fit simultaneously using the ``max\_model'' option of \texttt{SNooPy} \citep{burns11} to derive maximum-light magnitudes 
and the color-stretch parameters. A complete list of the SNe~Ia is found in Table~\ref{tab:CSP_SNe} of Appendix~\ref{sec:appendixA}, which 
includes the \sbv\ measurements, the observed ($B_{max} - V_{max}$) pseudocolors, and estimates of the host galaxy reddening, 
$E(B-V)_{\rm{host}}$, derived via an intrinsic color analysis like that described in detail by \cite{burns18}.
The maximum-light $uBVriYJH$ magnitudes may be accessed online at GitHub and Zenodo\footnote{\label{footnote4}GitHub: \url{https://github.com/syeduddin/fastdecliners}; 
Zenodo: \dataset[doi:10.5281/zenodo.17519375]{https://doi.org/10.5281/zenodo.17519375}.}.
These magnitudes have been 
corrected for Milky Way dust reddening using the \citet{schlafly11} recalibration of the \citet{schlegel98} infrared dust maps, and were 
K-corrected using the optical and near-infrared spectral energy templates from \citet{hsiao07} and \citet{lu23}, respectively, in combination 
with the CSP filter functions \citep{krisciunas17,phillips19}. 

For most of the analysis that follows, we define the subset of the CSP fast-declining SNe~Ia with redshifts $z_{\rm {CMB}} > 0.01$ as the
``Hubble Flow'' sample.  The second column of Table~\ref{tab:table1} gives the total number of SNe~Ia comprising this Hubble Flow sample for 
each of the CSP filters.

\begin{deluxetable}{ccc}[h]
\tabletypesize{\scriptsize}
\tablecolumns{3}
\tablewidth{0pt}
\tablecaption{Numbers of CSP Hubble Flow and IR~SBF calibrator SNe~Ia available in each filter\label{tab:table1}}
\tablehead{
\colhead{Filter} &
\colhead{Hubble Flow SNe~Ia\tablenotemark{a}} &
\colhead{IR SBF Calibrators\tablenotemark{b}}
}
\startdata
$u$ & 30 &  8  \\
$B$ & 43 & 12  \\
$g$ & 36 &  7  \\
$V$ & 43 & 12  \\
$r$ & 43 & 10  \\
$i$ & 42 & 10  \\
$Y$ & 36 &  7  \\
$J$ & 35 &  9  \\
$H$ & 29 &  9  \\
\enddata
\tablenotetext{a}{CSP-I and CSP-II SNe with \sbv~$< 0.75$ and $z_{\rm {CMB}} > 0.01$}
\tablenotetext{b}{SNe with NIR SBF distances \citep{jensen21,garnavich23} and \sbv~$< 0.75$}
\end{deluxetable}

\subsection{Distance Calibrators}
\label{sec:sample_calibrators}

As distance calibrators, we employ the SNe~Ia listed in Table~2 of \citet{garnavich23} with IR~SBF distances measured by \citet{jensen21} and  
\citet{garnavich23} and limited to color-stretch values of \sbv~$< 0.75$ as determined by \citet{uddin24}.  We also require that the SNe used
in the calibration have, at a minimum, $B$ and $V$ light curves that include coverage at maximum light.  The 12 SNe that meet these requirements 
are listed in Table~\ref{tab:SBF_SNe} of Appendix~\ref{sec:appendixB} along with their observed ($B_{max} - V_{max}$) pseudocolors, \sbv\ 
values,  $E(B-V)_{\rm{host}}$ estimates, and IR~SBF distance moduli.  The \texttt{SNooPy} max\_model magnitudes for these SNe~Ia are available online
at GitHub and Zenodo (see Footnote~\ref{footnote4}).  Like the CSP sample, all magnitudes and pseudocolors have been K-corrected as well as adjusted for 
Milky Way dust extinction.  The third column of Table~\ref{tab:table1} gives the number of IR~SBF distance calibrators available for each of 
the CSP filters.

\begin{figure*}[b]
\epsscale{.7}
\plotone{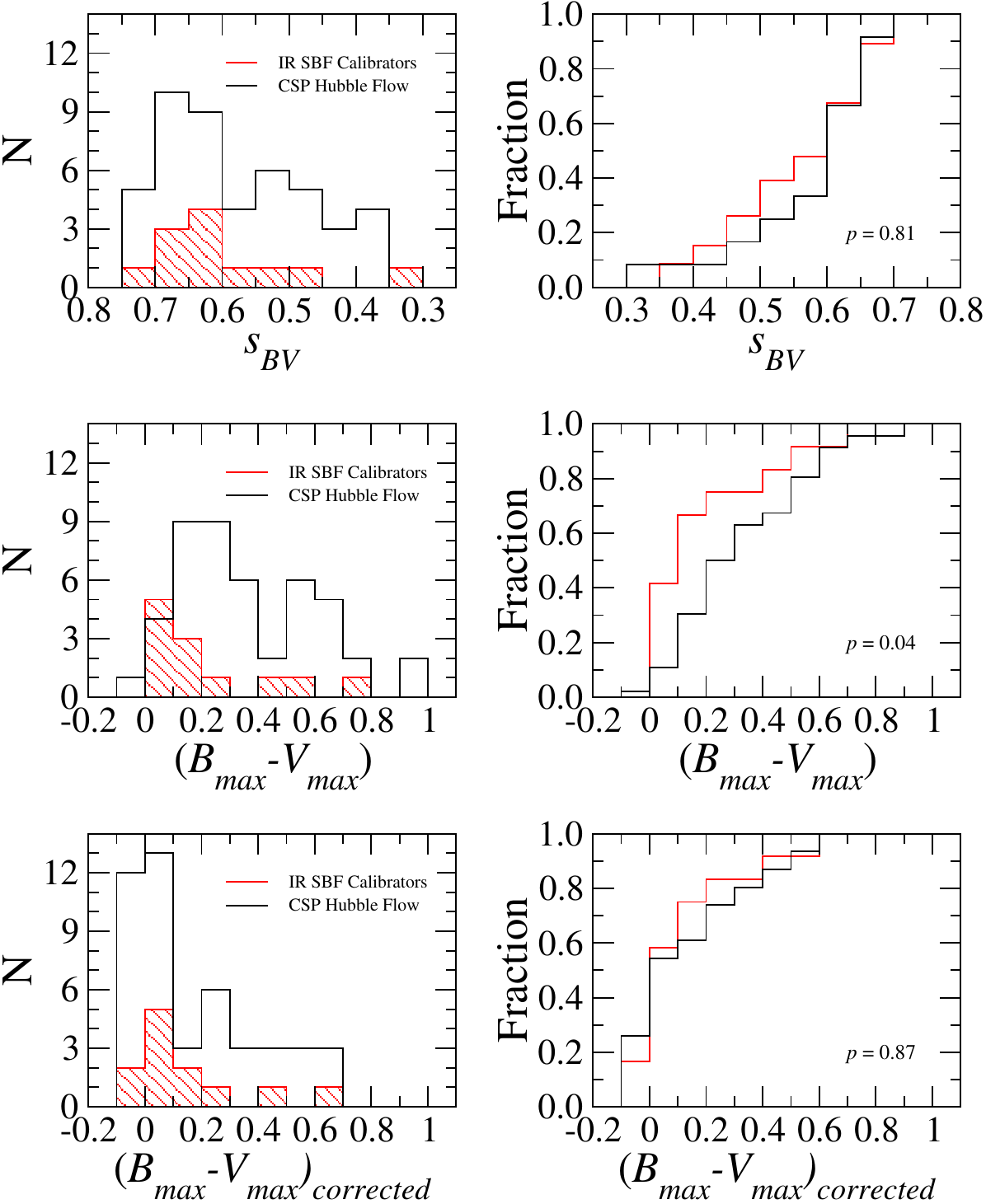} 
\caption{(upper row) Normal and cumulative histograms of the \sbv\ measurements of the CSP Hubble Flow ($z_{\rm {CMB}} > 0.01$) 
SNe~Ia and the IR~SBF calibrators (red hatched lines).  (middle row) Normal and cumulative histograms of the observed ($B_{max} - V_{max}$) 
pseudocolors for CSP Hubble Flow SNe~Ia and the IR~SBF calibrators (red hatched lines). (lower row) Normal and cumulative histograms of the 
($B_{max} - V_{max}$) pseudocolors, corrected for host dust reddening, for CSP Hubble Flow SNe~Ia and the IR~SBF calibrators (red hatched 
lines).  The K-S two-sample test $p$ values are given in the three cumulative histograms.}
\label{fig:hist_sBV_BV}
\end{figure*}

The top row of Figure~\ref{fig:hist_sBV_BV} shows normal and cumulative histograms of the \sbv\ values for the CSP Hubble Flow sample and the 
IR~SBF calibrators, and the middle row displays the normal and cumulative histograms of the observed ($B_{max} - V_{max}$) pseudocolors.
A two-sample Kolmogorov–Smirnov (K-S) test comparing the cumulative distribution of \sbv\ measurements of the Hubble Flow and IR~SBF 
calibrators gives a $p$-value of 0.81, indicating that there is no evidence that the two samples are drawn from different distributions at the
95\% confidence level.  However, a K-S test for these same two samples comparing the cumulative distribution of the observed 
($B_{max} - V_{max}$) pseudocolors returns a $p$-value of 0.04, implying that the hypothesis that the two samples come from the same 
distribution is rejected at the 95\% confidence level.  This is likely due to differences in the host galaxy reddenings between the two 
samples. Indeed, using the intrinsic color analysis technique of \cite{burns18}, we find a median color excess of the CSP Hubble Flow sample 
of $E(B-V)_{host} = 0.15$~mag, whereas for the IR~SBF calibrators it is 0.08~mag.  The bottom row of Figure~\ref{fig:hist_sBV_BV} shows normal 
and cumulative histograms of the ($B_{max} - V_{max}$) pseudocolors after correcting each SN individually for its host color excess.  
Repeating the K-S test gives a $p$-value of 0.87, confirming at the 95\% level that there is no evidence that the CSP Hubble Flow and IR~SBF 
calibrator samples are drawn from different distributions of intrinsic ($B_{max} - V_{max}$) values when differences in host reddening are 
accounted for\footnote{The topic of host galaxy dust extinction in fast-declining SNe~Ia will be discussed in a future paper on the general 
properties of these objects.}.

In a CSP study of the fast-declining IR~SBF calibrators SNe~2007on and 2011iv, both of which appeared in the same host galaxy (NGC~1404),
\citet{gall2018} found differences in the peak magnitudes
of $0.30 \pm 0.02$~mag and $0.20 \pm 0.01$~mag in the $B$ and $H$ bands, respectively,
after correcting for the dependence of luminosity on color and light curve shape.  These authors speculated that differences in the 
central density of the white dwarf progenitor just before ignition could lead to such a luminosity spread, and advised caution in using 
transitional SNe~Ia as cosmological standard candles.  As will be shown in \S\ref{sec:hubble}, the intrinsic scatter, $\sigma_{int}$, in peak 
absolute magnitude after correction for light curve shape and color is 0.17~mag in the $B$ band and 0.19~mag in $H$ for the full sample of 
CSP fast decliners and IR~SBF calibrators.  These values are very similar to the intrinsic dispersions for the full sample of CSP SNe~Ia 
using the same methodology \citep{uddin24}. The differences in magnitude for SNe~2007on and 2011iv are therefore equivalent to  
$\sim$1--2 times $\sigma_{int}$, which is within the observed scatter of the CSP Hubble Flow sample of fast decliners.  (The reader is 
referred to Appendix~\ref{sec:appendixC} for further discussion of SNe~2007on and 2011iv.)

\section{Analysis}
\label{sec:analysis}

In this section, we examine the absolute magnitudes of the CSP Hubble Flow sample of fast-declining SNe~Ia as a function of the color-stretch
parameter, \sbv, and separately, the observed ($B_{max} - V_{max}$) pseudocolor.  The absolute magnitudes were calculated from the host galaxy 
redshifts, $z_{\rm {CMB}}$, assuming $\Lambda$CDM cosmology with fixed Hubble constant $H_0 = 72$~\kmsmpc, mass density parameter
$\Omega_m = 0.27$, and cosmological constant parameter $\Omega_\Lambda = 0.73$.  The Hubble Flow sample is limited to those with 
$z_{\rm {CMB}} > 0.01$ to minimize the effects of peculiar velocities on the absolute magnitudes.  

\subsection{Absolute Magnitude versus \sbv}
\label{sec:M_vs_sBV}

\begin{figure*}[h]
\epsscale{.85}
\plotone{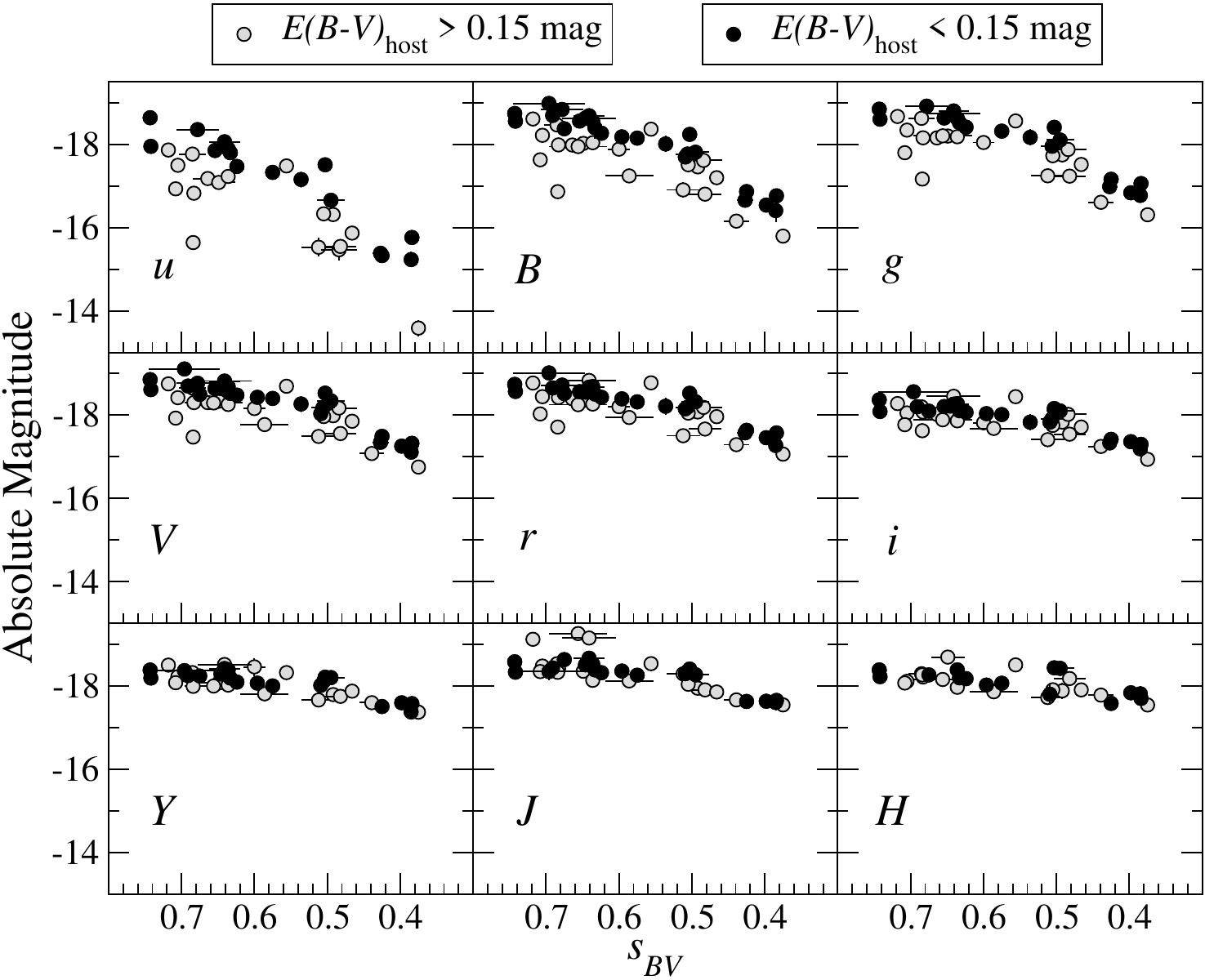}
\caption{Derived absolute magnitudes in $uBgVriYJH$ filters of the CSP SNe~Ia with $z_{\rm {CMB}} > 0.01$ plotted against \sbv. 
The data are corrected only for Milky Way extinction and the K-correction. The black circles correspond to the subset of SNe for 
which the host galaxy reddening, $E(B-V)\rm{_{host}}$ is estimated to be less than 0.15 mag (see text for further details).}
\label{fig:M_vs_sBV_all_filters}
\end{figure*}

Figure~\ref{fig:M_vs_sBV_all_filters} displays the dependence of the absolute magnitudes on the color-stretch parameter, \sbv, for the 
fast-declining Hubble Flow sample of SNe~Ia with no correction applied to the magnitudes other than for Milky Way extinction and the 
K-correction.  The correlation is the steepest and also has the highest dispersion in the bluest filters, and grows flatter and tighter in 
the reddest filters.  The fact that the dispersion in absolute magnitude decreases with increasing wavelength suggests that at least some of 
the scatter observed in the bluer filters may be due to uncorrected host galaxy extinction.  As argued by \citet{phillips12} and 
\citet{burns14}, the ``blue edge'' of the observed ($B_{max} - V_{max}$) versus \sbv\ relation can be used to estimate the amount of host 
galaxy extinction that SNe~Ia suffer. The black points in Figure~\ref{fig:BV_v_sBV} correspond to the fast-decliners for which we estimate 
that $E(B-V)_{\rm{host}} < 0.15$~mag.  These same SNe~Ia, also shown as black points in Figure~\ref{fig:M_vs_sBV_all_filters}, more tightly
delineate the absolute magnitude vs \sbv\ relationship for fast-decliners.  Second-order polynomial fits to these ``blue-edge'' SNe~Ia yield 
the rms dispersions given in Table~\ref{tab:table2}.  Apart from the $u$ filter, these values imply the possibility of measuring distances 
to 10\% or better if host reddening can be reliably corrected. Indeed, the dispersions in the $r$ through $J$ bands are remarkably low at 
$0.13$--$0.15$~mag without any corrections.  We suspect that the larger dispersion of $0.20$~mag in the $H$~band reflects the difficulty of 
making precise ground-based observations at this wavelength, rather than being intrinsic to SNe~Ia.  K-corrections are also more problematic 
in the $H$-band since the CSP did not employ a redder filter with which to ``anchor'' the color-matching (a.k.a. the ``mangling function'') 
of the template spectra to the photometry.

The reader will note that there are three points that lie \textit{above} the relationship in the $J$-band.  These correspond to SN2008bd, 
iPTF11ppn, and CSP15B (SN J13471211-2422171). The observations of SN~2008bd began $\sim$14~days after $B$ maximum, and there is a single 
$J$-band measurement at $+18$~days.  This is the worst light curve coverage of any of the fast decliners in the CSP sample.  iPTF11ppn is 
the second most-distant fast decliner in the sample, and was observed for a period of only $\sim$10 days. The three $J$-band observations 
obtained began 8~days after $B$ maximum.  The optical light curves of CSP15B have good temporal coverage, but only a single $J$-band 
measurement was obtained at $+25$~days.  There is reason to believe, therefore, that the displacements of the $J$-band absolute magnitudes 
for these three SNe reflect problems in extrapolating the observations to maximum light via the SNooPy $J$-template fit.  Nevertheless, 
since the photometry in the other CSP filters for these three SNe does not show similar offsets, we have no reason to exclude them from 
the $J$-band data.

\begin{deluxetable}{ccc}[h]
\tabletypesize{\scriptsize}
\tablecolumns{3}
\tablewidth{0pt}
\tablecaption{RMS Dispersion in Absolute Magnitude of Fast-declining ``Blue-Edge'' SNe~Ia\label{tab:table2}}
\tablehead{
\colhead{Filter} &
\colhead{RMS Dispersion (mag)\tablenotemark{a}} &
\colhead{Number of SNe~Ia}
}
\startdata
$u$ & 0.36 & 16 \\
$B$ & 0.19 & 25 \\
$g$ & 0.19 & 19 \\
$V$ & 0.18 & 24 \\
$r$ & 0.15 & 24 \\
$i$ & 0.15 & 23 \\
$Y$ & 0.14 & 19 \\
$J$ & 0.13 & 19 \\
$H$ & 0.20 & 15 \\
\enddata
\tablecomments{``Blue-Edge'' SNe~Ia are defined as having $E(B-V)_{\rm{host}} < 0.15$~mag}
\tablenotetext{a}{With respect to second order polynomial fits.}
\end{deluxetable}

\subsection{Absolute Magnitude versus ($B_{max} - V_{max}$)}
\label{sec:M_vs_BV}

\begin{figure*}[t]
\epsscale{.85}
\plotone{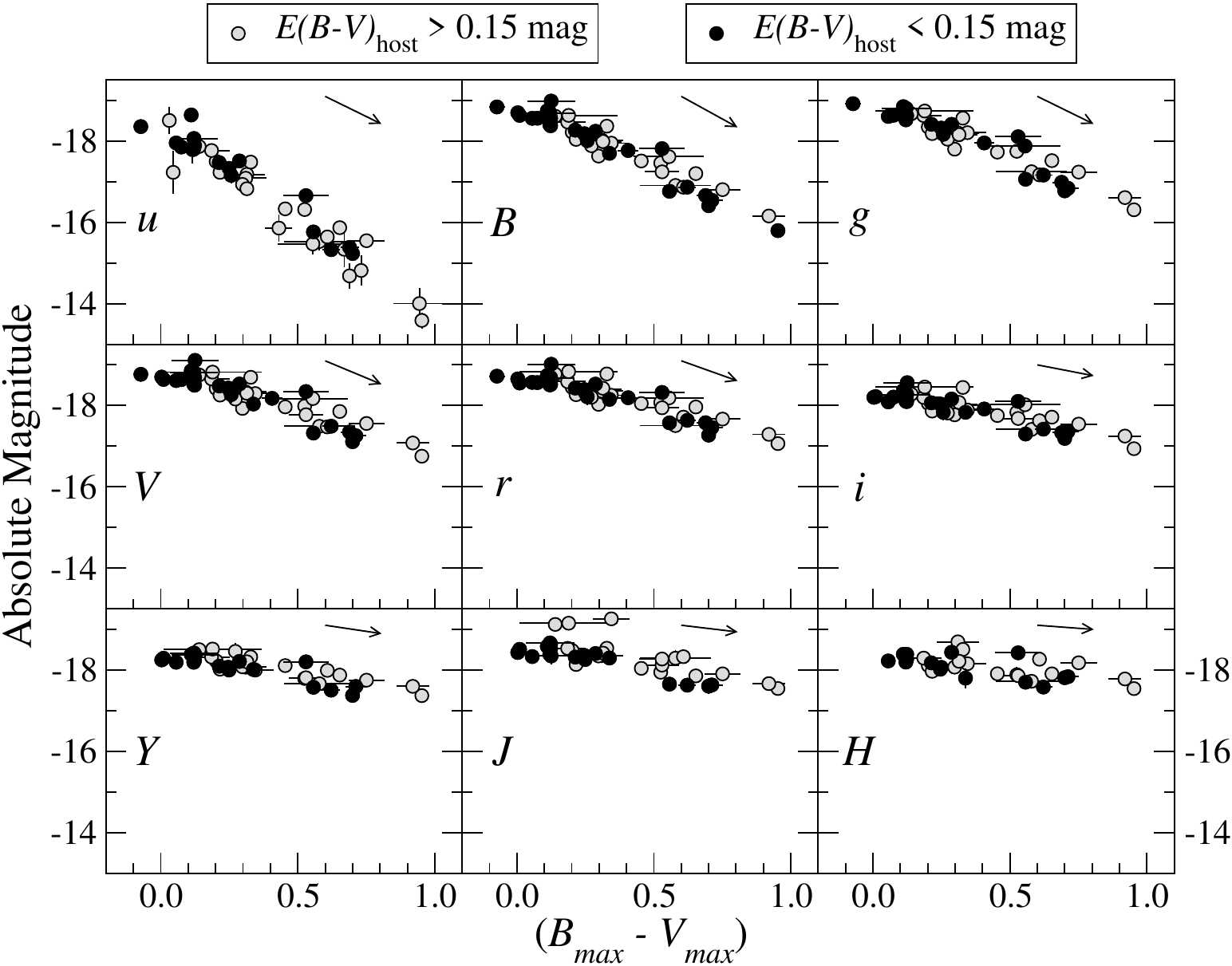}
\caption{Derived absolute magnitudes in $uBgVriYJH$ filters of the CSP SNe~Ia with $z_{\rm {CMB}} > 0.01$ plotted against the observed
($B_{max} - V_{max}$) pseudocolor.  \textit{The data are corrected only for Milky Way extinction and the K-correction.} The black circles 
correspond to the subset of SNe for which the host galaxy reddening, $E(B-V)\rm{_{host}}$, is estimated to be less than 0.15 mag (see text for 
further details). The arrows in each panel indicate the reddening vectors for host galaxy dust with a color excess of
$E(B-V)\rm{_{host}} = 0.2$~mag and $R_V = 2.8$.}
\label{fig:M_vs_B-V_all_filters}
\end{figure*}

Figure~\ref{fig:BV_v_sBV} shows that there is a strong correlation between \sbv\ and the observed ($B_{max} - V_{max}$) pseudocolor for SNe~Ia
with \sbv~$< 0.75$.  We therefore examine the dependence of absolute magnitude on observed ($B_{max} - V_{max}$) for the fast-decliners in
Figure~\ref{fig:M_vs_B-V_all_filters}.  In spite of the evidence presented in the previous section for significant host reddening for some SNe, 
strong correlations with surprisingly small dispersion are observed in all filters, including the $u$ band.  This unexpected result is due to a 
fortuitous coincidence between the slopes of the absolute magnitude versus observed ($B_{max} - V_{max}$) relations with the ratio of 
total-to-selective extinction values, $R_\lambda$, for each filter.  The arrows in each panel of Figure~\ref{fig:M_vs_B-V_all_filters} show 
the reddening vectors for dust with a color excess of $E(B-V) = 0.2$~mag and an $R_V$ value of 2.8, representative of values derived for normal, low-reddening SNe~Ia 
\citep[see, e.g.,][]{chotard11,phillips12,burns14,mandel17},
and Table~\ref{tab:table3} compares these values with linear polynomial fits to the data plotted in Figure~\ref{fig:BV_v_sBV}.  The coincidence
in slopes is not perfect, but close enough to significantly reduce the dispersion in absolute magnitude.

\begin{deluxetable}{cccc}[h]
\tabletypesize{\scriptsize}
\tablecolumns{4}
\tablecaption{Comparison of slopes of absolute magnitude versus observed ($B_{max} - V_{max}$) relations with total-to-selective extinction values and Tripp $\beta$ parameters\label{tab:table3}}
\tablehead{
\colhead{Filter} &
\colhead{\hspace{1.0cm}Slope\tablenotemark{a}}\hspace{1.0cm} &
\colhead{\hspace{0.5cm}$R_\lambda$\tablenotemark{b}}\hspace{0.5cm} &
\colhead{\hspace{0.5cm}$\beta$\tablenotemark{c}}\hspace{0.5cm}
}
\startdata
$u$  &  4.32 (08) & 4.4 & 4.12 (36) \\
$B$  &  2.79 (05) & 3.6 & 2.66 (22) \\
$g$  &  2.57 (06) & 3.3 & 2.42 (27) \\
$V$  &  1.82 (05) & 2.8 & 1.74 (21) \\
$r$  &  1.47 (05) & 2.4 & 1.39 (19) \\
$i$  &  1.13 (05) & 1.8 & 1.03 (20) \\
$Y$  &  0.87 (07) & 1.0 & 0.49 (18) \\
$J$  &  0.95 (07) & 0.8 & 0.30 (23) \\
$H$  &  0.62 (09) & 0.5 & 0.19 (27) \\
\enddata
\tablecomments{Errors are shown in parentheses.}
\tablenotetext{a}{Slope of absolute magnitude versus observed pseudocolor ($B_{max} - V_{max}$) as estimated via a linear polynomial fit to the 
  Hubble flow sample CSP SNe~Ia with \sbv~$< 0.75$.}
\tablenotetext{b}{The observed total-to-selective extinction values for dust with $R_V = 2.8$, calculated using the maximum-light 
 1991bg-like template spectrum, based on SN1991bg and SN1999by, of \citet[][\url{https://c3.lbl.gov/nugent/nugent_templates.html}]{nugent02}
 and the \citet{cardelli89} reddening law.}
 \tablenotetext{c}{Tripp method $\beta$ parameters from Table~\ref{tab:table4}.}
\end{deluxetable}

While there is no simple heuristic model that reproduces the luminosity-width relation, there is general agreement that the driving physics is
due to the amount and radial distribution of radioactive $^{56}$Ni produced due to the effects of the photon diffusion time and the efficiency 
of gamma-ray absorption \citep{nugent95,hoeflich95c,hoeflich96b,pinto00,shen21}.  Normal SNe~Ia around maximum light are characterized by 
($B_{max} - V_{max}$) colors near zero, as the dominant ionization stage changes from doubly- to singly-ionized species of iron-group elements
\citep{hoeflich95b}.   The photosphere adjusts to layers of varying ionization from doubly- to singly-ionized Fe-group elements.  Thus, the 
change in opacities by recombination determines the location of the photosphere, resulting in similar colors within different explosion 
scenarios \citep{hoeflich96,penney14,blondin2017}.  Instead, the diversity in intrinsic color is dominated by initial metallicity and white 
dwarf properties \citep{hoeflich17}.

In contrast, underluminous SNe~Ia tend to be redder and produce less $^{56}$Ni.  The apparent photosphere at maximum forms in the Si/S-rich 
region \citep{hoeflich96,hoeflich02,shen14} and the effective temperature is lower than for normal SNe~Ia \citep{nugent95}, leading to photospheres
dominated by singly-ionized elements.  For fast-declining SNe~Ia, the models show that the colors change quickly as a function of temperature.
which, in turn, is correlated with luminosity, and secondary effects are small.  Both the homogeneity and the dominance of a single ionization 
stage explain the robustness of the luminosity-color relation for fast decliners.

\section{The Hubble Constant}
\label{sec:hubble}

The ultimate test of whether fast-declining SNe~Ia can be employed to measure cosmological distances is to use these events to estimate
the Hubble constant, and then to compare the results with values obtained from samples of normal SNe~Ia.  In this section, we use the CSP Hubble 
Flow fast-decliners (\S\ref{sec:sample_CSP}) in combination with the IR~SBF calibrators (\S\ref{sec:sample_calibrators}) to infer values of 
the Hubble constant via the standard \citet{tripp98} method. In addition, we introduce a second method based solely on the strong correlation
between absolute magnitude and the observed ($B_{max} - V_{max}$) pseudocolor highlighted in \S\ref{sec:M_vs_BV}.  We choose to use the IR~SBF 
technique for calibration since it offers the largest number of uniform distance measurements to fast decliners. The Python notebooks used to calculate the Hubble constant in the section are available at GitHub (see Footnote~\ref{footnote4}).

\subsection{Tripp Method}
\label{sec:tripp}

Our analysis utilizes the same methodology employed by \citet{uddin24}.  The Tripp method incorporates both a correction for the light-curve 
shape, \sbv, and the maximum-light color, for which we use the observed pseudocolor ($B_{max} - V_{max}$).  A third correction for a host 
galaxy ``mass step'' \citep{kelly07,Lampeitl10,sullivan10} is often also included, but we ignore this correction for two reasons. First of all,
the fast-declining SNe~Ia occur almost exclusively in early-type galaxies and thus represent a more homogeneous sample of hosts compared to the
mixture of active and passive star-forming galaxies that typically comprise the samples of normal SNe~Ia used to determine $H_0$.  Secondly, 
our own work based on a sample of more than 300 SNe~Ia in the local universe with precise CSP photometry suggests that the mass step is small 
or, perhaps, even consistent with zero \citep{uddin24}.  We consider the possible systematic error of neglecting this correction in 
\S\ref{sec:systematic}.

Following Equation~(1) of \citet{uddin24}, we define the distance moduli, $\mu_{Tripp}$, of our sample of Hubble Flow SNe~Ia and
IR~SBF calibrators as:

\begin{equation}\label{eq:mu_obs1}
    \mu_{Tripp} = m_x -P^N_x(s_{BV}-0.5) -\beta_x(B_{max}-V_{max}-0.4).
\end{equation}

\noindent Here, $m_x$ is the peak apparent magnitude in filter $x$, $P^N_x(s_{BV} - 0.5)$ is a polynomial of order $N$ as a function of 
($s_{BV} - 0.5)$, and $\beta_x$ is the slope of the luminosity–-color relation.  We subtract $0.5$ from \sbv\ in 
order to minimize errors in the $P^N_x$ terms since this value coincides approximately with the middle of the range of \sbv\ values of our 
Hubble Flow sample.
Likewise, the mean pseudocolor of $0.4$~mag of the sample is subtracted from 
($B_{max}-V_{max})$.

Separately, we also define ``model distance moduli,'' $\mu_{model}$, for both the Hubble Flow and the IR~SBF calibrators.  For the Hubble Flow 
SNe~Ia, $\mu_{model}$ is calculated from both the heliocentric and CMB-frame redshifts in the $\Lambda$CDM framework as follows:

\begin{eqnarray}\label{eq:mulambdacdm}
  \mu_{\Lambda CDM}  = 5\log_{10} & & \bigg[\bigg(\frac{1+z_{hel}}{1+z_{cmb}} \bigg)\frac{cz_{cmb}}{H_0}\bigg(1+ 
  \frac{1-q_0}{2}z_{cmb}\bigg)\bigg] +25.
\end{eqnarray}

\noindent In this equation, $H_0$ is a free parameter and $q_0 = \Omega_M/2 - \Omega_\Lambda$ is fixed at a value of –0.538 \citep{planck16}.
For the calibrators, $\mu_{model}$ is taken to be the IR~SBF distance moduli.  That is, 

\begin{equation}\label{eq:mumodchoice}
    \mu_{model} =
    \begin{cases}
      \mu_{\Lambda CDM} & \rm Hubble \ Flow \ SNe \ Ia \\
      \mu_{IR \ SBF} & \rm IR \ SBF \ calibrators
    \end{cases}
\end{equation}

We compare the $\mu_{Tripp}$ and $\mu_{model}$ values by defining a $\chi^2$ as:

\begin{equation}\label{chi}
    \chi^2 =  \sum_{i} \frac{(\mu_{Tripp,i}-\mu_{model,i})^2}{\sigma^2_i+\sigma^2_{int}+\sigma^2_{pec}}.
\end{equation}

\noindent Here, $\sigma^2_i$ is the sum of the individual variances of the observed quantities along with the covariance between peak 
magnitude and color and the covariance between peak magnitude and the color-stretch parameter. Equation~(5) of \citet{uddin24} gives the expression for $N =2$.
The term, $\sigma^2_{pec}$ is meant to account for the uncertainty due to the peculiar velocities, $v_{pec}$, of the host galaxies.
\citet{uddin24} left $\sigma^2_{pec}$ as a free parameter, but we fix it at 300~\kms, which is appropriate for a Hubble flow sample
at $z_{\rm{CMB}} > 0.01$\footnote{Fixing $v_{pec}$ from values of 200--350~\kms\ does not significantly change our results for
$H_0$ or $\sigma^2_{int}$.}.  Finally, $\sigma^2_{int}$ is the intrinsic random scatter of the entire sample of Hubble Flow SNe~Ia and 
IR~SBF calibrators, which we solve for as a free parameter.

The most likely values and statistical uncertainties for $P^N$, $\beta$, $\sigma_{int}$, and $H_0$ for each filter are calculated via Markov 
Chain Monte Carlo (MCMC) techniques as explained in detail by \citet{uddin24}.  
The results across all filters are given 
in Table~\ref{tab:table4} where solutions for both $N = 1$ (linear) and $N = 2$ (quadratic) for the 
$P^N_x(s_{BV} - 0.5)$ term are compared. The priors for these solutions are specified
in Appendix~\ref{sec:appendixD}.  
As is seen, the results for the $N = 1$ and $N = 2$ cases differ only slightly in terms of
the individual $H_0$ and $\sigma^2_i$ values for each filter, and so we adopt the $N = 1$ solutions. 
A corner plot of the posterior probability distributions of variables from the output of the 
linear ($N = 1$) Tripp method solution for the $B$ filter is shown in Figure~\ref{fig:Tripp_corner_plot} of Appendix~\ref{sec:appendixD}.
Averaging the values of $H_0$ weighted by the statistical errors gives $75.5$~\kmsmpc.  Since the measurements in 
each filter are not fully independent, we adopt as the uncertainty a value of 3.1~\kmsmpc, which is the average of the nine statistical errors.

\begin{deluxetable*}{lccccccc}
\tabletypesize{\scriptsize}
\tablecolumns{8}
\tablewidth{0pt}
\tablecaption{H$_0$ and Nusiance Parameters for Tripp and Color Methods\label{tab:table4}}
\tablehead{
\colhead{Method} &
\colhead{Filter} &
\colhead{H$_0$} &
\colhead{$\sigma_{int}$} &
\colhead{P0} &
\colhead{P1} &
\colhead{P2} &
\colhead{$\beta$}
}
\startdata
                   & $u$ & 77.69 (4.57) & 0.26 (0.04) & $-16.47$ (0.11) &  $-0.28$ (0.77) &  \nodata & 4.53 (0.37) \\
                   & $B$ & 75.05 (2.76) & 0.19 (0.03) & $-17.52$ (0.07) &  $-0.59$ (0.55) &  \nodata & 2.88 (0.25) \\
                   & $g$ & 78.89 (3.67) & 0.21 (0.03) & $-17.67$ (0.09) &  $-0.32$ (0.63) &  \nodata & 2.65 (0.30) \\
Tripp              & $V$ & 75.65 (2.71) & 0.20 (0.03) & $-17.92$ (0.07) &  $-0.58$ (0.55) &  \nodata & 1.91 (0.24) \\
($N = 1$)          & $r$ & 75.31 (2.53) & 0.17 (0.02) & $-17.98$ (0.07) &  $-0.70$ (0.48) &  \nodata & 1.51 (0.20) \\
                   & $i$ & 76.12 (2.48) & 0.17 (0.02) & $-17.68$ (0.06) &  $-0.64$ (0.44) &  \nodata & 1.17 (0.19) \\
                   & $Y$ & 72.60 (2.35) & 0.13 (0.02) & $-17.90$ (0.06) &  $-1.03$ (0.40) &  \nodata & 0.72 (0.18) \\
                   & $J$ & 77.22 (3.21) & 0.19 (0.03) & $-17.95$ (0.08) &  $-1.79$ (0.61) &  \nodata & 0.66 (0.27) \\
                   & $H$ & 75.06 (3.38) & 0.19 (0.03) & $-17.93$ (0.08) &  $-1.05$ (0.60) &  \nodata & 0.46 (0.30) \\ 
\hline
                   & $u$ & 78.03 (4.13) & 0.25 (0.04) & $-16.48$ (0.10) &   4.61 (0.21) &  \nodata & \nodata     \\
                   & $B$ & 75.95 (2.71) & 0.19 (0.03) & $-17.54$ (0.07) &   3.10 (0.12) &  \nodata & \nodata     \\
                   & $g$ & 78.81 (3.55) & 0.20 (0.03) & $-17.69$ (0.09) &   2.78 (0.15) &  \nodata & \nodata     \\
Color              & $V$ & 75.92 (2.57) & 0.19 (0.03) & $-17.94$ (0.07) &   2.12 (0.13) &  \nodata & \nodata     \\
($N = 1$)          & $r$ & 76.06 (2.50) & 0.17 (0.02) & $-18.01$ (0.07) &   1.74 (0.11) &  \nodata & \nodata     \\
                   & $i$ & 76.88 (2.45) & 0.17 (0.02) & $-17.70$ (0.06) &   1.40 (0.12) &  \nodata & \nodata     \\
                   & $Y$ & 74.44 (2.47) & 0.14 (0.03) & $-17.92$ (0.07) &   1.09 (0.11) &  \nodata & \nodata     \\
                   & $J$ & 80.75 (3.45) & 0.22 (0.03) & $-17.97$ (0.09) &   1.34 (0.15) &  \nodata & \nodata     \\
                   & $H$ & 77.55 (3.26) & 0.20 (0.03) & $-17.94$ (0.08) &   0.88 (0.16) &  \nodata & \nodata     \\
\hline
\hline
                   & $u$ & 76.78 (4.29) & 0.26 (0.04) & $-16.52$ (0.11) &  $-1.87$ (0.99) &  7.33 (4.31) & 4.12 (0.36) \\
                   & $B$ & 74.10 (2.40) & 0.16 (0.02) & $-17.62$ (0.07) &  $-2.10$ (0.64) &  9.66 (2.68) & 2.66 (0.22) \\
                   & $g$ & 77.39 (3.34) & 0.18 (0.03) & $-17.80$ (0.09) &  $-1.90$ (0.74) & 10.42 (2.94) & 2.42 (0.27) \\
Tripp              & $V$ & 74.59 (2.42) & 0.18 (0.03) & $-18.03$ (0.07) &  $-2.03$ (0.62) & 10.36 (2.77) & 1.74 (0.21) \\
($N = 2$)          & $r$ & 74.57 (2.48) & 0.16 (0.02) & $-18.06$ (0.07) &  $-1.71$ (0.55) &  7.20 (2.53) & 1.39 (0.19) \\
                   & $i$ & 75.41 (2.38) & 0.15 (0.02) & $-17.75$ (0.07) &  $-1.70$ (0.59) &  6.94 (2.60) & 1.03 (0.20) \\
                   & $Y$ & 72.32 (2.27) & 0.13 (0.03) & $-17.93$ (0.06) &  $-2.08$ (0.51) &  4.99 (2.25) & 0.49 (0.18) \\
                   & $J$ & 76.25 (2.92) & 0.17 (0.03) & $-18.00$ (0.08) &  $-3.21$ (0.64) &  6.45 (2.78) & 0.30 (0.22) \\
                   & $H$ & 74.09 (3.31) & 0.19 (0.03) & $-17.96$ (0.08) &  $-2.02$ (0.74) &  4.46 (3.21) & 0.19 (0.27) \\ 
\hline
                   & $u$ & 76.94 (4.01) & 0.25 (0.04) & $-16.61$ (0.13) &   4.60 (0.21) &  1.67 (0.79) & \nodata     \\
                   & $B$ & 74.81 (2.53) & 0.17 (0.03) & $-17.63$ (0.08) &   3.15 (0.13) &  1.07 (0.52) & \nodata     \\
                   & $g$ & 78.48 (3.56) & 0.19 (0.03) & $-17.75$ (0.11) &   2.77 (0.15) &  0.69 (0.64) & \nodata     \\
Color              & $V$ & 75.10 (2.57) & 0.19 (0.03) & $-18.03$ (0.08) &   2.15 (0.13) &  0.98 (0.53) & \nodata     \\
($N = 2$)          & $r$ & 75.72 (2.60) & 0.17 (0.02) & $-18.06$ (0.08) &   1.78 (0.12) &  0.72 (0.45) & \nodata     \\
                   & $i$ & 76.58 (2.57) & 0.16 (0.02) & $-17.74$ (0.08) &   1.39 (0.12) &  0.47 (0.48) & \nodata     \\
                   & $Y$ & 75.59 (2.62) & 0.16 (0.03) & $-17.92$ (0.08) &   1.05 (0.12) &  0.49 (0.49) & \nodata     \\
                   & $J$ & 82.16 (3.98) & 0.24 (0.04) & $-17.98$ (0.12) &   1.28 (0.17) &  0.61 (0.69) & \nodata     \\
                   & $H$ & 78.90 (3.86) & 0.21 (0.03) & $-17.90$ (0.12) &   0.84 (0.18) & $-0.05$ (0.68) & \nodata     \\
\enddata
\tablecomments{Errors are shown in parentheses.}
\end{deluxetable*}

\subsection{Color Method}

The tight correlation between absolute magnitude and observed pseudocolor for the fast-decliners presented in \S\ref{sec:M_vs_BV}
suggests that the \sbv\ term in Equation~(\ref{eq:mu_obs1}) can be dispensed with.  That is, the distance moduli can be written
more simply as:

\begin{equation}\label{eq:mu_obs2}
    \mu_{Color} = m_x -P^N_x(B_{max}-V_{max} - 0.4).
\end{equation}

\noindent In this case, $P^N_x(B_{max}-V_{max} - 0.4)$ is a polynomial of order $N$ as a function of ($B_{max}-V_{max}-0.4$).  We subtract the approximate mean pseudocolor of our Hubble Flow sample of $0.4$~mag from 
($B_{max}-V_{max})$ in order to minimize errors in the $P^N$ terms.  For $\mu_{model}$, we adopt the same definition given in 
Equation~(\ref{eq:mumodchoice}).
Note that in this ``Color'' method, the expression for $\sigma^2_i$ takes a slightly different form than Equation~(5) of \citet{uddin24},
since the $s_{BV}$ term and associated covariances are not required. Ignoring the index, for
each SN~Ia we define $\sigma^2_i$ in the $N=2$ case as:

\begin{equation}
 \begin{split}
        \sigma^2 = & \sigma_{m_x}^2 + (P1 + 2P2)\ [(B-V)-0.4)]^2\sigma^2_{(B-V)}\\
         &  - 2(P1 + 2P2)\ [(B-V)-0.4)]\ cov(m_x,(B-V)).
    \end{split}   
\end{equation}

The parameters derived from the MCMC analysis for this Color method for solutions with
$N =1$ and $N = 2$ are given in Table~\ref{tab:table4}.  As for the Tripp method, the results differ very little.  A corner plot of the posterior probability distributions of variables from the output of the 
linear ($N = 1$) Color method solution for the $B$ filter is shown in Figure~\ref{fig:Color_corner_plot} of Appendix~\ref{sec:appendixD}.
Averaging the values for all 
filters in the linear case yields $H_0 = 76.7$ \kmsmpc.  This value is in excellent agreement with that obtained using the Tripp method in \S\ref{sec:tripp}.
Averaging the nine statistical errors of the measurements for each filter gives an uncertainty of $3.0$~\kmsmpc\ that is also comparable.

\subsection{Systematic Errors}
\label{sec:systematic}

The calibration of the \citet{jensen21} and \citet{garnavich23} IR~SBF distances is tied to Cepheid observations of the LMC and 16 spiral 
galaxies in the Virgo and Fornax clusters \citep{garnavich23}, and hence is anchored to the Cepheid distance scale.  According to 
\citet{blakeslee21}, the zero point of the IR~SBF distances has a 3.1\% systematic error that must be added to our $H_0$ error 
values\footnote{Recently, \cite{jensen25} presented an alternative calibration of the IR~SBF distances using a geometrical calibration of the 
TRGB method applied to nearby elliptical galaxies that, in turn, can be used to measure Cepheid-free IR~SBF distances.
However, our aim is to compare our results with those of \citet{uddin24}, which used the \citet{blakeslee21} calibration.}.

The fact that we have ignored a host-galaxy mass correction in our calculations is another possible source of systematic error.  Using the 
values given in \citet{uddin24}, the median host mass of the Hubble Flow sample is $log_{10}(M_{host}/M_\odot) = 10.7$, while for the 
calibrators it is $log_{10}(M_{host}/M_\odot) = 11.5$.  Both values are on the high side of the ``mass step,'' which typically is placed at a 
value of $log_{10}(M_{host}/M_\odot) \sim $10.0--10.3.  Hence, if the host mass correction is actually a step, we are justified in ignoring it.
However, it is possible that the correction should, more properly, be modeled as a continuous linear function of host mass.  Under this 
assumption, \citet{uddin24} found a typical slope of $-0.4~\rm{mag~dex}^{-1}$.  If we ignore this correction, the calibrators will be 
systematically $\sim0.03$~mag brighter than the Hubble Flow sample after light curve shape and color corrections.  So the distance moduli that 
we derive for the Hubble Flow sample will be $\sim0.03$~mag larger, resulting in an $H_0$ that is 1.4\% lower.  This translates to a possible 
systematic error of 1.0~\kmsmpc, which we are justified in ignoring considering the relatively large statistical errors of our $H_0$ 
measurements.

To summarize, we find that the Tripp method yields a Hubble constant of $H_0 = 75.5 \pm 3.1~\rm{(statistical)}^{+2.5}_{-2.3}~\rm{(systematic)}$~\kmsmpc, 
while the Color method gives $H_0 = 76.7 \pm 3.0~\rm{(statistical)}^{+2.6}_{-2.4}~\rm{(systematic)}$~\kmsmpc, where the systematic errors
correspond to the Cepheid zero point error and the possible mass step error added in quadrature.  We emphasize that these are the 
\textit{minimum} systematic errors in our $H_0$ calculations.  For example, the fact that the approximate slopes of the luminosity-color 
relations are slightly different from typical host galaxy reddening vectors may introduce some additional systematic error in the
Color method.  It is also conceivable that the dust that reddens fast-declining SNe~Ia in early-type galaxies could have significantly 
different properties than what is observed for normal SNe~Ia.  However, modeling such effects is beyond the scope of this paper.

\section{Interpreting the Tripp Method Parameters}
\label{sec:tripp_parameters}

\begin{figure*}[t]
\epsscale{.5}
\plotone{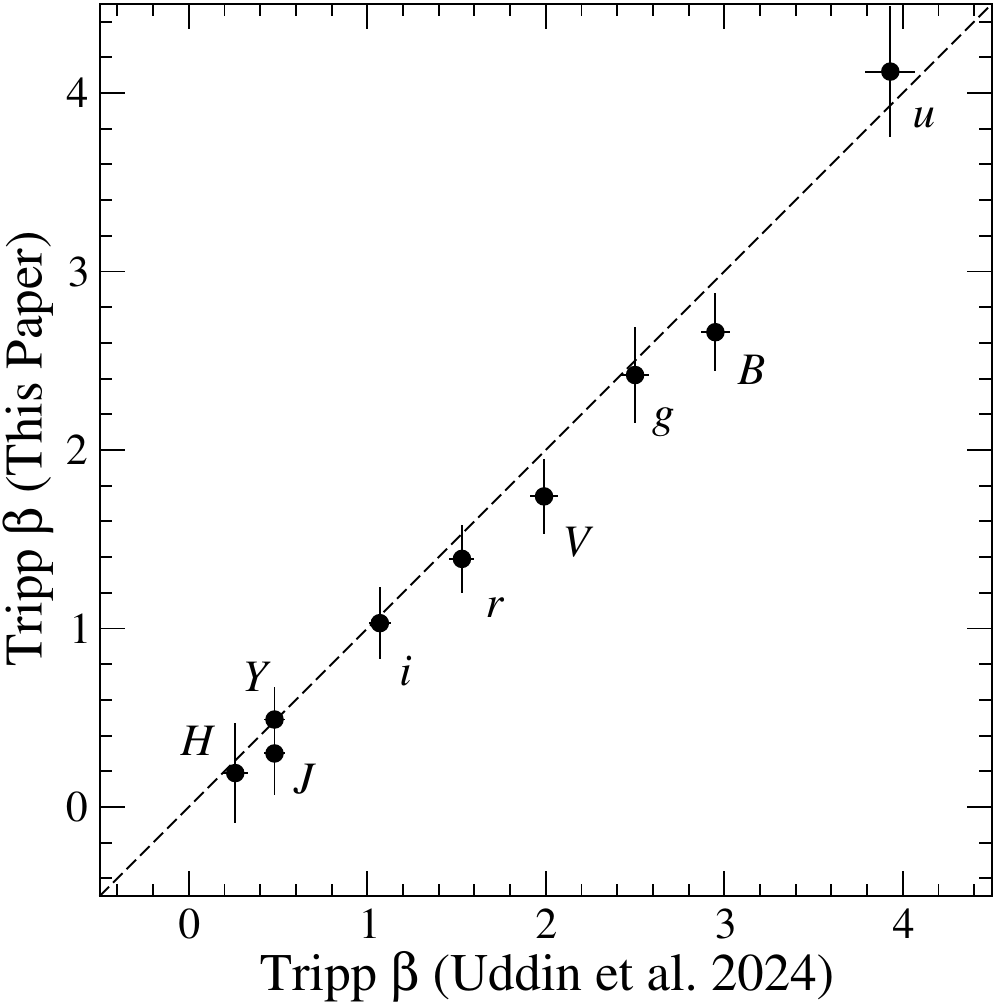}
\caption{Comparison of $\beta$ values for the Tripp method derived by \citet{uddin24} for a sample of $>300$~SNe~Ia spanning the full
range of luminosity and \sbv\ plotted against the values determined for the subset of fast-decliners in the present paper.  Each point is 
labeled by the filter to which it corresponds.
In both cases, a quadratic ($N = 2$) luminosity term is assumed.}
\label{fig:beta}
\end{figure*}

The Tripp method formula is conventionally thought of as consisting of a luminosity correction that depends on light curve shape, and a second 
term to correct for color.  However, both intrinsic (i.e., SNe~Ia explosion physics) and extrinsic (dust extinction) sources contribute to the 
observed colors of SNe~Ia.  The $\beta$ parameter of the Tripp method combines these two sources into a single parameter, which, as emphasized by 
\citet{mandel17} and \citet{burns18}, is conceptually an oversimplification.  As mentioned previously, for normal SNe~Ia with \sbv~$> 0.75$ 
(\dm~$\lesssim 1.5$~mag), the dependence of the observed ($B_{max} - V_{max}$) pseudocolor on light curve shape is relatively weak, but for fast 
decliners there is a strong intrinsic component (see Figure~\ref{fig:BV_v_sBV}). Naively, therefore, we might expect the derived $\beta$ 
parameters for our sample of fast decliners to differ significantly from those found for a sample of normal SNe~Ia.  In Figure~\ref{fig:beta}, 
we compare the $\beta$ values derived in the present paper with the values found by \citet{uddin24} for the full CSP sample of $\sim$300 
SNe~Ia, for which events with \sbv~$< 0.75$ make up less than 20\% of the total number.  Somewhat surprisingly, it is seen that the $\beta$ 
values are actually in agreement at the $\sim1\sigma$ level.

\begin{figure*}[t]
\epsscale{1.0}
\plotone{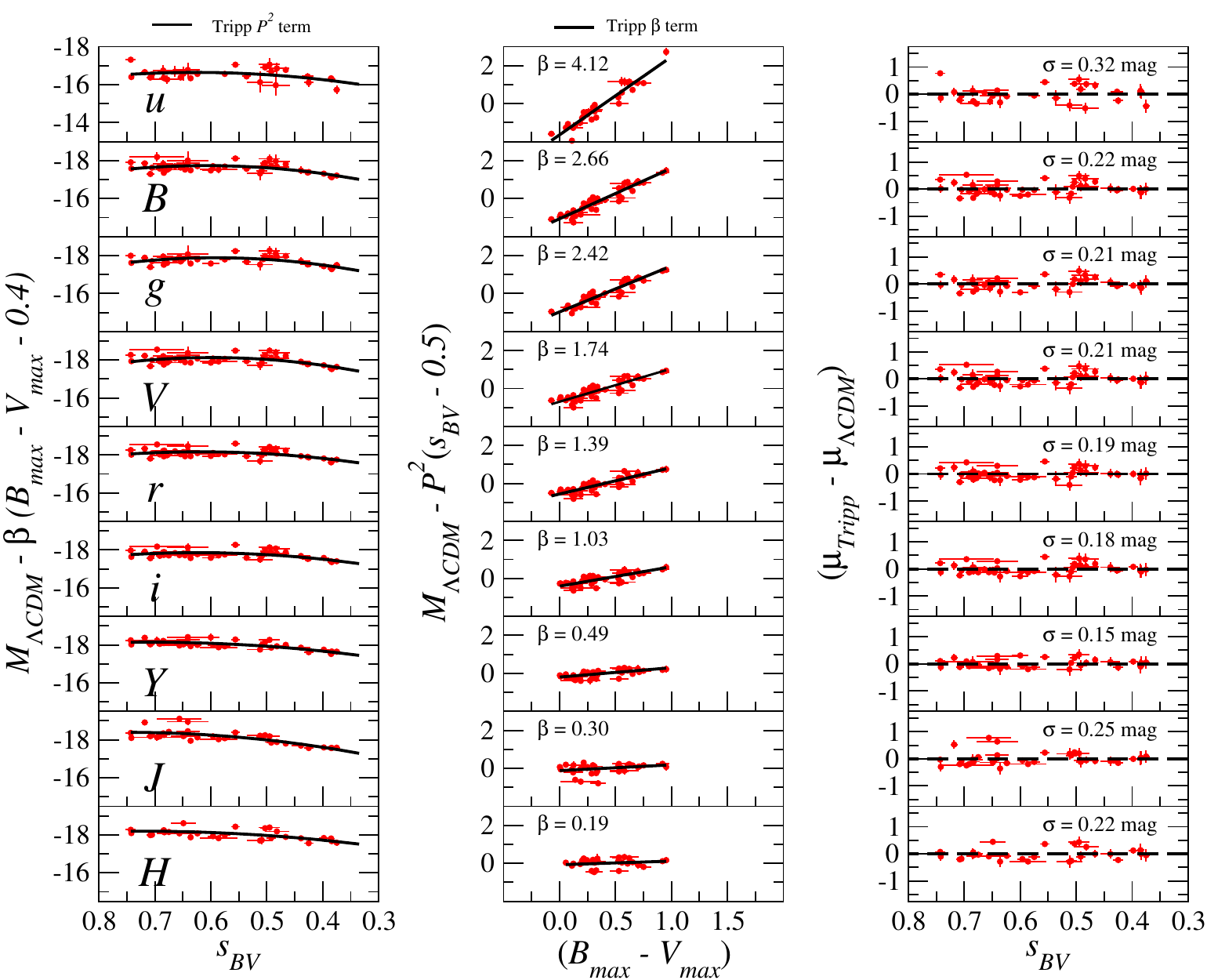}
\caption{(left column) Observed absolute magnitudes, $M_{\Lambda CDM}$, in $uBgVriYJH$ filters, for the Hubble Flow sample of CSP 
fast-declining SNe~Ia with \sbv\ $< 0.75$ and $z_{\rm {CMB}} > 0.01$, calculated from Equation~(\ref{eq:Mlambdacdm}) using the $H_0$ values
from Table~\ref{tab:table4}, and plotted against \sbv. The $\beta (B_{max}-V_{max}-0.4)$ color term of Equation~(\ref{eq:mu_obs1}) has been 
subtracted from the magnitudes, with the black lines showing the fits to the Tripp method polynomials, $P^2(s_{BV} - 0.5)$ for each 
filter. (middle column) Residual magnitudes for the Hubble Flow sample after subtraction of the $P^2(s_{BV} - 0.5)$ luminosity correction of
Equation~(\ref{eq:mu_obs1}) from the observed absolute magnitudes, $M_{\Lambda CDM}$.  Here the results are plotted against the observed 
($B_{max} - V_{max}$) pseudocolors, with the black lines corresponding to the fits of the Tripp $\beta$ parameters from Table~\ref{tab:table4},
which are also listed in each panel.  (right column) Hubble residuals, ($\mu_{Tripp}-\mu_{\Lambda CDM}$), for the Hubble Flow sample plotted 
against \sbv\ after fully correcting for both the $P^2(s_{BV} - 0.5)$ and $\beta (B_{max}-V_{max}-0.4)$ terms.  RMS dispersions are listed for 
each filter.}
\label{fig:Tripp_fast_decliners}
\end{figure*}

To understand why this is the case, we must examine in detail how (and why) the Tripp method actually works by comparing the MCMC fits
to the observations of the Hubble Flow sample of fast-declining SNe~Ia.
For this comparison, we assume a quadratic ($N = 2$) luminosity term.
From Equation~(\ref{eq:mulambdacdm}), we calculate observed 
absolute magnitudes at maximum in the $\Lambda$CDM framework for each SN as:

\begin{eqnarray}\label{eq:Mlambdacdm}
  M_{\Lambda CDM} = m_{max} - \mu_{\Lambda CDM},
\end{eqnarray}

\noindent where $m_{max}$ is the apparent maximum-light magnitude.  In the left-hand column of plots in Figure~\ref{fig:Tripp_fast_decliners}, 
these absolute magnitudes are plotted versus \sbv\ after subtracting off the ``color'' terms, $\beta (B_{max}-V_{max}-0.4)$, of 
Equation~(\ref{eq:mu_obs1}), with the $\beta$ values for each filter taken from Table~\ref{tab:table4}.
Overplotted are the fits of the ``luminosity'' terms, $P^2(s_{BV} - 0.5)$, found by the MCMC analysis and calculated from the parameters 
in Table~\ref{tab:table4}.  If we compare these Tripp luminosity terms with the actual luminosity-width relations shown in 
Figure~\ref{fig:M_vs_sBV_all_filters}, they are significantly flatter, especially for the bluest filters.  This is because of the strong 
dependence of luminosity on color at these wavelengths (see Figure~\ref{fig:M_vs_B-V_all_filters}), which has been removed by subtraction 
of the $\beta (B_{max}-V_{max}-0.4)$ term.

In the middle column of Figure~\ref{fig:Tripp_fast_decliners}, the absolute magnitudes, $M_{\Lambda CDM}$, are again plotted, but this time 
versus the observed ($B_{max} - V_{max}$) pseudocolors and after subtracting off the luminosity terms, $P^2(s_{BV} - 0.5)$, of 
Equation~(\ref{eq:mu_obs1}), with the $P^2$ parameters taken from Table~\ref{tab:table4}.  The slopes of the linear fits to the data 
correspond to the Tripp $\beta$ parameters for each filter.  As shown in Table~\ref{tab:table3}, these $\beta$ values are similar to the 
approximate slopes that we measured for the absolute magnitude versus ($B_{max} - V_{max}$) pseudocolor relations.  \textit{In other words, 
for the fast decliners, these $\beta$ parameters largely reflect the intrinsic temperature variation that is driven by the declining 
production of $^{56}$Ni with decreasing luminosity at these faster decline rates.}

\begin{figure*}[t]
\epsscale{1.0}
\plotone{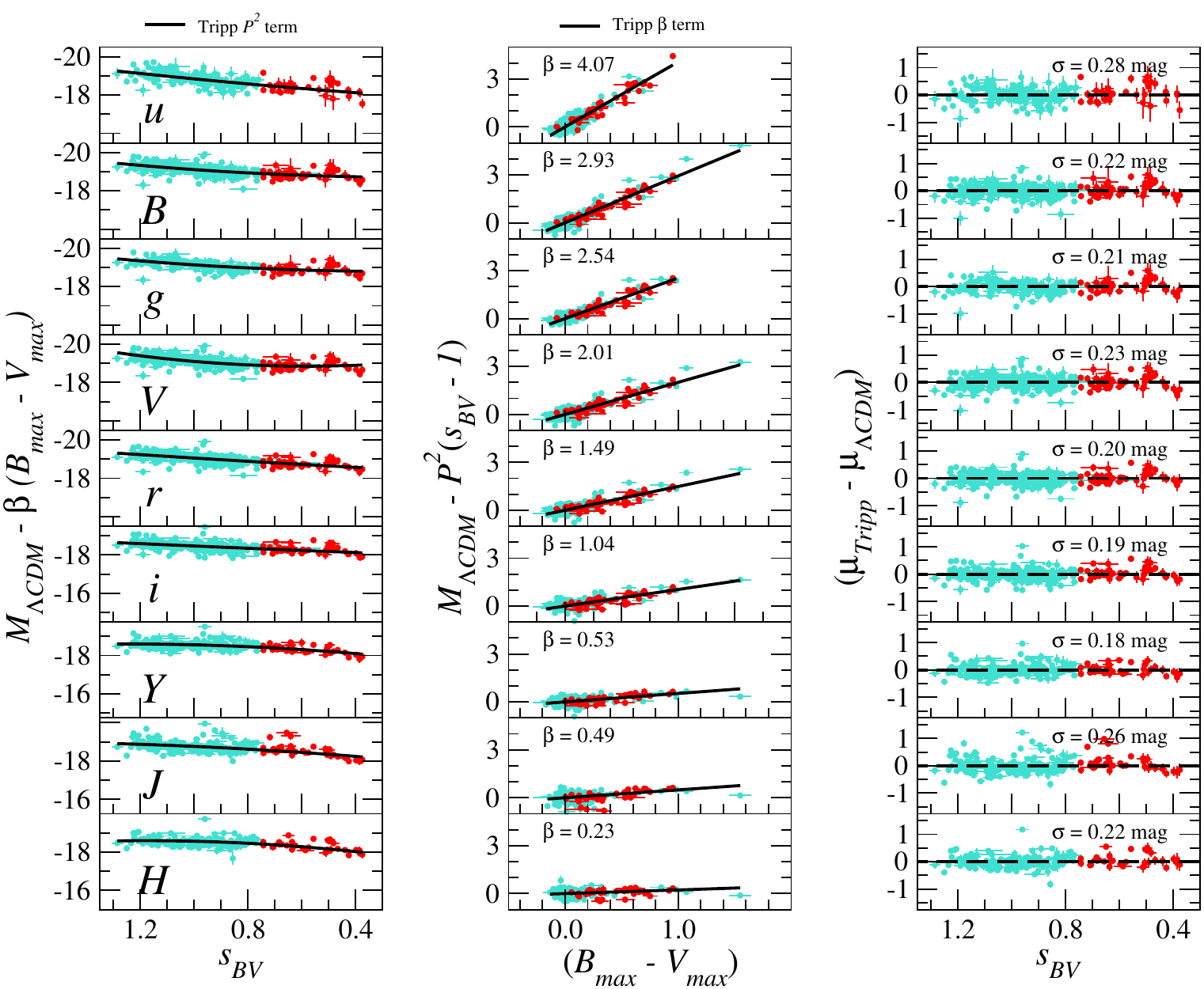}
\caption{Same as Figure~\ref{fig:Tripp_fast_decliners}, but showing the full sample of $\sim$300 CSP SNe~Ia from \citet{uddin24} with 
$z_{\rm {CMB}} > 0.01$.  The SNe are divided into normal ones with \sbv\ $> 0.75$ (turquoise points) and fast decliners with \sbv\ $< 0.75$ 
(red points).  Note that, due to the much larger range of \sbv\ values in the \citet{uddin24}
sample, these authors subtracted 1.0 from \sbv\ to minimize errors in the $P^2$ 
luminosity term.}
\label{fig:Tripp_Uddin}
\end{figure*}

Figure~\ref{fig:Tripp_Uddin} shows analogous plots for a Tripp method solution for the full sample of $\sim$300 CSP SNe~Ia with 
$z_{\rm {CMB}} > 0.01$ analyzed by \citet{uddin24}.  Here the SNe are divided into those having normal (\sbv\ $> 0.75$; turquoise points) or 
fast-declining (\sbv\ $< 0.75$; red points) light curves.  The left-hand column of plots in this figure again displays the observed absolute
magnitudes plotted against \sbv\ after subtraction of the color terms using the parameters of the MCMC analysis 
for $z_{\rm {CMB}} > 0.01$ given in Table~11 of \citet{uddin24}.  In the middle column of plots, the luminosity terms have been 
subtracted from the observed absolute magnitudes and plotted versus color.  Here we see the key result that the residuals for 
the normal (\sbv\ $> 0.75$) SNe~Ia, which are mostly due to host galaxy dust extinction, spread out in color and display in 
($B_{max} - V_{max}$) very similar slopes to the residuals for the fast decliners (\sbv\ $< 0.75$).  On the other hand, the fast-decliners 
spread out in color due mostly to temperature, and this slope is remarkably similar to the slope of the normal SNe~Ia from dust. \textit{It 
is exactly this coincidence that makes possible the Color method  discussed in \S\ref{sec:M_vs_BV}.}  Indeed, it is this coincidence that
also enables the Tripp method to give reliable $H_0$ measurements for the full CSP sample which includes both normal and fast-declining SNe~Ia.

In the right-hand columns of plots of Figures~\ref{fig:Tripp_fast_decliners} and \ref{fig:Tripp_Uddin}, the Hubble residuals, 
($\mu_{Tripp}~-~\mu_{\Lambda CDM}$), calculated after application of both the {luminosity and color terms are 
displayed along with the RMS dispersions.  The very similar dispersions obtained for the full CSP sample of SNe~Ia 
(Figure~\ref{fig:Tripp_Uddin}) versus the restricted sample of events with \sbv\ $< 0.75$ (Figure~\ref{fig:Tripp_fast_decliners}) clearly 
demonstrate that a fast decliner can be used to measure cosmological distances with the same precision as normal SNe~Ia.

\section{Discussion and Conclusions}
\label{sec:conclusions}

Figure~\ref{fig:H0} displays our Hubble constant results as a function of filter for the two methods employed in \S\ref{sec:hubble}.  The 
individual values are consistent across the range of filters from the ultraviolet to the near-infrared although, not surprisingly, the filters
with the most calibrators ($BVri$) show less dispersion in their values than do the filters with fewer calibrators ($ugYJH$).  Using the 
\citet{jensen21} and \citet{garnavich23} IR SBF distance moduli, we find that the Tripp method yields a value of 
$H_0 = 75.5 \pm 3.1~\rm{(statistical)}^{+2.5}_{-2.3}~\rm{(systematic)}$~\kmsmpc, while the Color method gives 
$H_0 = 76.7 \pm 3.0~\rm{(statistical)}^{+2.6}_{-2.4}~\rm{(systematic)}$~\kmsmpc.  Both are in excellent agreement with the Tripp method 
result of \citet{uddin24} of $H_0 = 76.3 \pm 1.3~\rm{(statistical)}$~\kmsmpc\ (weighted average of all filters, adopting the average of the 
errors of the individual filter measurements as the uncertainty) for the full sample of $\sim$300 CSP SNe~Ia using the IR~SBF calibrators.  
To be fair, nine of the 22 IR~SBF calibrators used by \citet{uddin24} were fast decliners with \sbv~$> 0.75$, and so the calculations are not
completely independent.  Nevertheless, this outcome confirms that fast-declining SNe~Ia can, by themselves, serve as excellent cosmological 
standard candles when \sbv\ is used as the light-curve width parameter, anchored by a set of local calibrators of choice --- Cepheids in our 
case here.

\begin{figure*}[t]
\epsscale{1.}
\plotone{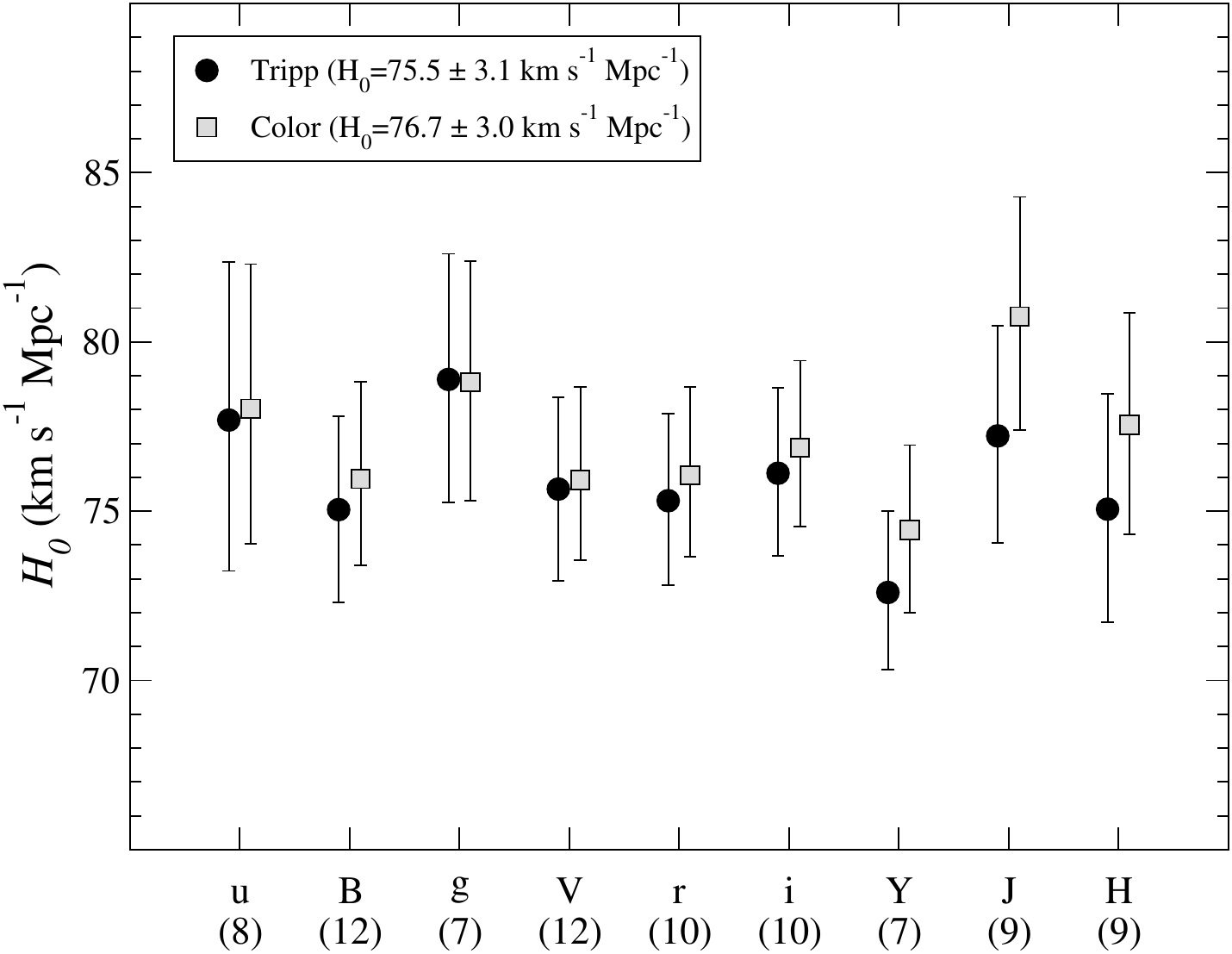}
\caption{Comparison of $H_0$ values for the CSP filters using the linear ($N = 1$) solutions for the Tripp and Color methods.  The values are based on the \citet{jensen21} and
\citet{garnavich23} distance moduli for the IR~SBF calibrators, the numbers of which are given in parentheses on the x-axis for each filter.}
\label{fig:H0}
\end{figure*}
 
By far the most surprising finding of this study is that fast-declining SNe~Ia {can provide an estimate $H_0$ using only their observed 
($B_{max} - V_{max}$) pseudocolors and no light curve shape information.  This means that SNe~Ia identified as fast-decliners via, for example, 
an optical spectrum, and whose maximum light magnitudes in $B$ and $V$ are known, {can be used to measure distances to their host galaxies
without the necessity of measuring \sbv\ or knowing anything about the dust extinction produced by the host.  \textit{This is a property that is 
unique to fast-declining SNe~Ia.}

The Tripp method $\beta$ parameter is often interpreted as being related to host galaxy dust reddening. However, we find that for the fast 
decliners, the value of $\beta$ coincides with the slope of the absolute magnitude versus ($B_{max} - V_{max}$) pseudocolor relation, which, 
in turn, is related to the dependence of peak luminosity on temperature.  It is the close coincidence of the $\beta$ values of the fast 
decliners with the ratio of total-to-selective extinction values, $R_\lambda$, for dust that makes possible the Color method.

Fast-declining SNe~Ia have largely been ignored by previous studies due to their faintness with respect to normal events.  However, as seen in 
Figure~4 of \citet{burns18}, the difference at maximum light in the NIR is $\lesssim1$~mag for all but the fastest decliners.  Of course, since 
fast-declining SNe~Ia are associated with older stellar populations, their true usefulness as distance indicators is in the local universe since 
they will become increasingly rare at larger lookback times.  For now, the number of fast-declining SNe~Ia with independently-measured host 
galaxy distances is small, which accounts for the relatively large statistical errors of our Hubble constant calculation.  But this situation 
should improve significantly over the next few years thanks to JWST.  It will be interesting to incorporate TRGB and PNLF distances to 
host galaxies as more of these become available, since this will allow a better estimate of the systematic error in using fast-declining 
SNe-Ia to infer $H_0$.  Recent JWST TRGB measurements to four early-type host galaxies by \citet{newman25} are an important first step in this 
direction.  More high-quality light curves of fast-declining SNe~Ia in the Hubble flow are also needed, which facilities such as the Rubin 
Observatory should eventually provide.  \textit{We therefore encourage observers not to ignore fast-declining SNe Ia in the future as these 
objects provide an important additional tool for measuring cosmological distances.}

\begin{acknowledgments}
The work of the CSP-I and CSP-II was generously supported by the National Science Foundation under 
grants AST0306969, AST0607438, AST1008343, AST-1613426, AST-1613455, and AST-1613472.
The CSP-II was also supported in part by the Danish Agency for Science and Technology and Innovation through a 
Sapere Aude Level 2 grant.
M.S. is funded by the Independent Research Fund Denmark (IRFD, grant number  10.46540/2032-00022B).
C.A. and E.B. acknowledge support from NASA grants JWST-GO-02114,
JWST-GO-02122, JWST-GO-04522, JWST-GO-04217, JWST-GO-04436,
JWST-GO-03726, JWST-GO-05057, JWST-GO-05290, JWST-GO-06023,
JWST-GO-06677, JWST-GO-06213, JWST-GO-06583. Support for
programs \#2114, \#2122, \#3726, \#4217, \#4436, \#4522,  \#5057,
\#6023, \#6213, \#6583, and \#6677
were provided by NASA through a grant from the Space Telescope Science
Institute, which is operated by the Association of Universities for Research in Astronomy, Inc., under NASA contract NAS 5-03127.
E.B. is supported in part by NASA grant 80NSSC20K0538.
L.G. acknowledges financial support from AGAUR, CSIC, MCIN and AEI 10.13039/501100011033 under projects PID2023-151307NB-I00, 
PIE 20215AT016, CEX2020-001058-M, ILINK23001, COOPB2304, and 2021-SGR-01270.
KK, NBS, and SU thank the George P. and Cynthia Woods Mitchell Institute for Fundamental Physics and Astronomy for financial suppport.
We gratefully acknowledge the use of WISeREP -- https://wiserep.weizmann.ac.il.
This research has made use of the \citet{ned19},
which is funded by the National 
Aeronautics and Space Administration and operated by the California Institute of Technology.
Finally, we acknowledge the Las Campanas Observatory for the outstanding support during our observing runs and the Carnegie Observatories 
Time Allocation Committee for generous time allocations.
\end{acknowledgments}

\vspace{5mm}
\facilities{Magellan:Baade (FourStar wide-field near-infrared camera), 
Swope (SITe3 CCD imager, e2v 4K x 4K CCD imager),
du~Pont (SITe2 CCD imager, Tek5 CCD imager, RetroCam near-infrared imager), 
La Silla-QUEST, CRTS, PTF, iPTF, OGLE, ASAS-SN, KISS)}

\software{SNooPy \citep{burns11}}

\pagebreak

\appendix

\restartappendixnumbering

\section{Fast-Declining CSP SNe~Ia}
\label{sec:appendixA}

The following table lists the 54 fast-declining (\sbv~$< 0.75$) SNe~Ia observed by the CSP. The SN names, \sbv, and observed 
($B_{max} - V_{max}$) values are taken from \citet{uddin24}.  The maximum-light $uBgVriYJH$ magnitudes for each SN, likewise taken from 
\citet{uddin24}, are available online (see Footnote~\ref{footnote4}).

\begin{longrotatetable}
\begin{deluxetable*}{llcccccl}
\tabletypesize{\scriptsize}
\tablecolumns{8}
\tablewidth{0pt}
\tablecaption{Fast-Declining SNe~Ia Observed by the CSP\label{tab:CSP_SNe}}
\tablehead{
\colhead{SN} &
\colhead{Host Galaxy} &
\colhead{$z_{CMB}$} &
\colhead{\sbv} &
\colhead{($B_{max} - V_{max}$)\tablenotemark{a}} &
\colhead{$E(B-V)_{host}$\tablenotemark{b}} &
\colhead{Sample} &
\colhead{Filters Observed}
}
\startdata
2004gs  &  MCG +03-22-020                 & 0.0283 & 0.705 (0.003) &  0.202 (0.006) & 0.23 (0.02) &  CSP-I  & $u,B,g,V,r,i,Y,J,H$ \\
2005bl  &  NGC 4059                       & 0.0251 & 0.427 (0.011) &  0.689 (0.034) & 0.19 (0.03) &  CSP-I  & $u,B,g,V,r,i$       \\
2005ke  &  NGC 1371                       & 0.0045 & 0.438 (0.003) &  0.670 (0.012) & 0.18 (0.02) &  CSP-I  & $u,B,g,V,r,i,Y,J,H$ \\
2005mc  &  UGC 4414                       & 0.0260 & 0.664 (0.020) &  0.314 (0.067) & 0.30 (0.03) &  CSP-I  & $u,B,g,V,r,i$       \\
2006bd  &  UGC 6609                       & 0.0268 & 0.398 (0.009) &  0.712 (0.039) & 0.14 (0.03) &  CSP-I  & $B,g,V,r,i,Y,J,H$   \\
2006eq  &  2MASX J21283758+0113490        & 0.0484 & 0.649 (0.023) &  0.310 (0.077) & 0.27 (0.03) &  CSP-I  & $u,B,g,V,r,i,Y,J,H$ \\
2006gj  &  UGC 2650                       & 0.0270 & 0.708 (0.007) &  0.298 (0.012) & 0.35 (0.02) &  CSP-I  & $u,B,g,V,r,i,Y,J,H$ \\
2006gt  &  2MASX J00561810-013732         & 0.0436 & 0.575 (0.010) &  0.249 (0.016) & 0.05 (0.02) &  CSP-I  & $u,B,g,V,r,i,Y,J,H$ \\
2006hb  &  MCG -04-12-34                  & 0.0153 & 0.685 (0.018) &  0.185 (0.066) & 0.15 (0.03) &  CSP-I  & $u,B,g,V,r,i,Y,J,H$ \\
2006mr  &  NGC1316                        & 0.0055 & 0.335 (0.004) &  0.731 (0.028) & 0.03 (0.02) &  CSP-I  & $u,B,g,V,r,i,Y,J,H$ \\
2006ob  &  UGC 1333                       & 0.0583 & 0.743 (0.009) &  0.111 (0.015) & 0.12 (0.01) &  CSP-I  & $u,B,g,V,r,i,Y,J,H$ \\
2007N   &  MCG -01-33-012                 & 0.0140 & 0.375 (0.006) &  0.953 (0.024) & 0.33 (0.03) &  CSP-I  & $u,B,g,V,r,i,Y,J,H$ \\
2007al  &  2MASX J09591870-1928233        & 0.0133 & 0.385 (0.006) &  0.700 (0.025) & 0.09 (0.02) &  CSP-I  & $u,B,g,V,r,i,Y,J,H$ \\
2007ax  &  NGC 2577                       & 0.0076 & 0.404 (0.008) &  0.689 (0.026) & 0.20 (0.03) &  CSP-I  & $u,B,g,V,r,i,Y,H$   \\
2007ba  &  UGC 9798                       & 0.0391 & 0.556 (0.008) &  0.328 (0.019) & 0.18 (0.02) &  CSP-I  & $u,B,g,V,r,i,Y,J,H$ \\
2007hj  &  NGC 7461                       & 0.0129 & 0.636 (0.005) &  0.216 (0.009) & 0.16 (0.02) &  CSP-I  & $u,B,g,V,r,i,Y,J,H$ \\
2007jh  &  CGCG 391-014                   & 0.0403 & 0.600 (0.015) &  0.272 (0.026) & 0.18 (0.03) &  CSP-I  & $B,g,V,r,i,Y$       \\
2007mm  &  SHK 035                        & 0.0689 & 0.506 (0.016) &  0.406 (0.037) & 0.15 (0.03) &  CSP-I  & $B,g,v,r,i$         \\
2007ol  &  2MASX J01372378-0018422        & 0.0549 & 0.678 (0.029) & -0.073 (0.029) & 0.01 (0.01) &  CSP-I  & $u,B,g,V,r$         \\
2007on  &  NGC 1404                       & 0.0062 & 0.588 (0.003) &  0.116 (0.012) & 0.01 (0.01) &  CSP-I  & $u,B,g,V,r,i,Y,J,H$ \\
2007ux  &  2MASX J10091969+1459268        & 0.0320 & 0.624 (0.006) &  0.213 (0.010) & 0.13 (0.02) &  CSP-I  & $u,B,g,V,r,i,Y,J,H$ \\
2008O   &  ESO 256-G11                    & 0.0393 & 0.683 (0.010) &  0.314 (0.016) & 0.36 (0.02) &  CSP-I  & $u,B,g,V,r,i,Y,J,H$ \\
2008R   &  NGC 1200                       & 0.0129 & 0.633 (0.006) &  0.120 (0.011) & 0.08 (0.02) &  CSP-I  & $u,B,g,V,r,i,Y,J,H$ \\
2008bd  &  MCG -02-26-42                  & 0.0313 & 0.656 (0.040) &  0.345 (0.068) & 0.27 (0.03) &  CSP-I  & $B,g,V,r,i,Y,J,H$   \\
2008bi  &  NGC 2618                       & 0.0144 & 0.512 (0.023) &  0.579 (0.129) & 0.31 (0.04) &  CSP-I  & $u,B,g,V,r,i,Y,J,H$ \\
2008bt  &  NGC 3404                       & 0.0166 & 0.492 (0.008) &  0.525 (0.013) & 0.18 (0.02) &  CSP-I  & $u,B,g,V,r,i,Y,J,H$ \\
2009F   &  NGC 1725                       & 0.0129 & 0.384 (0.007) &  0.557 (0.026) & 0.01 (0.02) &  CSP-I  & $u,B,g,V,r,i,Y,J,H$ \\
ASAS15ga  &  NGC 4866                       & 0.0077 & 0.496 (0.030) &  0.431 (0.050) & 0.22 (0.04) &  CSP-II & $u,B,g,V,r,i,Y,J,H$ \\
CSP12J\tablenotemark{c}    &  2MASX J06070178-6921180        & 0.0149 & 0.484 (0.025) & 0.192 (0.064) &  0.56 (0.13)  &  CSP-II & $u,B,g,V,r,i$       \\
CSP13aao\tablenotemark{c}  &  2MASX J05583036-6333386        & 0.0616 & 0.696 (0.049) & 0.046 (0.029) &  0.13 (0.09)  &  CSP-II & $B,V,r,i,Y$         \\
CSP15B\tablenotemark{c}    &  ESO 509-G108                   & 0.0208 & 0.718 (0.006) & 0.178 (0.018) &  0.14 (0.01)  &  CSP-II & $u,B,g,V,r,i,Y,J$   \\
CSP15aae\tablenotemark{c}  &  NGC 5490                       & 0.0170 & 0.505 (0.004) & 0.163 (0.018) &  0.45 (0.01)  &  CSP-II & $u,B,g,V,r,i,Y,J,H$ \\
KISS15m   &  NGC 4098                       & 0.0254 & 0.425 (0.009) &  0.622 (0.031) & 0.10 (0.03) &  CSP-II & $u,B,g,V,r,i,Y,J,H$ \\
LSQ11pn\tablenotemark{c}   &  2MASX J05164149+0629376        & 0.0327 & 0.503 (0.008) & 0.016 (0.013) &  0.29 (0.02)  &  CSP-II & $u,B,g,V,r,i,Y,J,H$ \\
LSQ12fvl  &  MCG -06-12-002                 & 0.0560 & 0.596 (0.009) &  0.246 (0.017) & 0.14 (0.02) &  CSP-II & $B,V,r,i,Y,J,H$     \\
LSQ13dkp  &  2MASX J03101094-3638017        & 0.0685 & 0.646 (0.009) &  0.010 (0.013) & 0.02 (0.01) &  CSP-II & $B,V,r,i,Y$         \\
LSQ14ip   &  2MASX J09442084+0435319        & 0.0624 & 0.509 (0.010) &  0.338 (0.023) & 0.13 (0.02) &  CSP-II & $B,V,r,i,Y,J,H$     \\
LSQ14jp   &  2MASX J12572166-1547411        & 0.0465 & 0.675 (0.007) &  0.123 (0.010) & 0.14 (0.02) &  CSP-II & $B,V,r,i,Y,J,H$     \\
LSQ14act  &  2MASX J15594429-1026396        & 0.0595 & 0.691 (0.007) &  0.003 (0.010) & 0.05 (0.02) &  CSP-II & $B,V,r,i,Y$         \\
LSQ14ajn\tablenotemark{c}  &  CGCG 068-091                   & 0.0222 & 0.654 (0.004) & 0.075 (0.009) & 0.05 (0.01) &  CSP-II & $u,B,g,V,r,i$       \\
LSQ14gfb  &  2MASX J05100559-3618388        & 0.0528 & 0.586 (0.033) &  0.530 (0.062) & 0.44 (0.04) &  CSP-II & $B,V,r,i,Y,J,H$     \\
PTF11ppn  &  2MASX J21352164+2656051        & 0.0662 & 0.641 (0.037) &  0.189 (0.179) & 0.24 (0.05) &  CSP-II & $B,V,r,i,Y,J,H$     \\
PTF11pra\tablenotemark{c}  &  NGC 881                        & 0.0167 & 0.439 (0.017) & 0.920 (0.060) & 0.44 (0.04) & CSP-II & $B,g,V,r,i,Y,J,H$   \\
iPTF13ebh &  NGC 0890                       & 0.0125 & 0.636 (0.004) &  0.123 (0.009) & 0.08 (0.02) &  CSP-II & $u,B,g,V,r,i,Y,J,H$ \\
iPTF14w   &  UGC 07034                      & 0.0201 & 0.742 (0.004) &  0.056 (0.005) & 0.09 (0.01) &  CSP-II & $u,B,g,V,r,i,Y,J,H$ \\
iPTF14aje &  SDSS J152512.43-014840.1       & 0.0282 & 0.684 (0.005) &  0.608 (0.011) & 0.65 (0.03) &  CSP-II & $u,B,g,V,r,i,Y,J,H$ \\
2011iv  &  NGC 1404                       & 0.0060 & 0.699 (0.007) &  0.031 (0.011) & 0.07 (0.01) &  CSP-II & $u,B,g,V,r,i,Y,J,H$ \\
2011jn  &  2MASX J12571157-1724344        & 0.0485 & 0.641 (0.021) &  0.121 (0.088) & 0.07 (0.03) &  CSP-II & $u,B,g,V,r,i,Y,J$   \\
2012ij  &  CGCG 097-050                   & 0.0121 & 0.536 (0.008) &  0.257 (0.016) & 0.02 (0.01) &  CSP-II & $u,B,g,V,r,i$       \\
2013ay  &  IC 4745                        & 0.0157 & 0.495 (0.019) &  0.530 (0.083) & 0.13 (0.03) &  CSP-II & $u,B,g,V,r,i,Y,J,H$ \\
2013bc  &  IC 4209                        & 0.0235 & 0.482 (0.022) &  0.751 (0.066) & 0.34 (0.04) &  CSP-II & $u,B,g,V,r,i,Y,J,H$ \\
2014ba  &  NGC 7410                       & 0.005  & 0.338 (0.020) &  0.944 (0.094) & 0.24 (0.04) &  CSP-II & $u,B,g,V,r,i,Y,J,H$ \\
2014dn  &  IC 2060                        & 0.0221 & 0.466 (0.005) &  0.653 (0.010) & 0.23 (0.02) &  CSP-II & $u,B,g,V,r,i,Y,J,H$ \\
2015bp\tablenotemark{c} &  NGC 5839                       & 0.0047 &  0.693 (0.003) &  0.045 (0.007) & 0.07 (0.01) &  CSP-II & $u,B,g,V,r,i,Y,J,H$ \\
\enddata
\tablecomments{Errors are shown in parentheses.}
\tablenotetext{a}{Values are K-corrected and corrected for Milky Way reddening. No correction has been applied for host galaxy reddening.}
\tablenotetext{b}{Estimate of the host galaxy reddening, $E(B-V)_{host}$, derived via the intrinsic color analysis described in detail by \cite{burns18}}
\tablenotetext{c}{CSP12J = OGLE-2012-SN-040; CSP13aao = OGLE-2013-SN-123; CSP15B = PSN J13471211-2422171; CSP15aae = CSS150214:140955+1703155 (2015bo);
                  LSQ11pn = 2011jq; LSQ14ajn = 2014ah; PTF11pra = 2011hk; 2015bp = SNhunt281}.
\end{deluxetable*}
\end{longrotatetable}

\clearpage

\restartappendixnumbering

\section{IR~SBF Calibrators}
\label{sec:appendixB}

The following table lists the 12 fast-declining (\sbv~$< 0.75$) SNe~Ia used as IR~SBF calibrators in this paper. The SN names, \sbv, and 
observed ($B_{max} - V_{max}$) values are taken from \citet{uddin24}.  The maximum-light $uBgVriYJH$ magnitudes for each SN, likewise taken 
from \citet{uddin24}, are available online (see Footnote~\ref{footnote4}).  The IR~SBF distance moduli for SN~2006mr, SN~2007on, and SN~2011iv 
are from \citet{garnavich23}; those for the remaining SNe are from \citet{jensen21}

\begin{longrotatetable}
\begin{deluxetable*}{llccccclc}
\tabletypesize{\scriptsize}
\tablecolumns{9}
\tablewidth{0pt}
\tablecaption{Fast-Declining SNe~Ia with IR SBF Distances\label{tab:SBF_SNe}}
\tablehead{
\colhead{SN} &
\colhead{Host Galaxy} &
\colhead{$z_{CMB}$} &
\colhead{\sbv} &
\colhead{($B_{max} - V_{max}$)\tablenotemark{a}} &
\colhead{$E(B-V)_{host}$\tablenotemark{b}} &
\colhead{$\mu_{NIR~SBF}$} &
\colhead{Filters Observed} &
\colhead{Photometry\tablenotemark{c}}
}
\startdata
2006mr  &  NGC1316                        & 0.0055 & 0.335 (0.004) &  0.731 (0.028)  & 0.03 (0.02) & 31.200 (0.093) & $u,B,g,V,r,i,Y,J,H$ & 1 \\
2007cv  &  IC 2597                        & 0.0087 & 0.711 (0.010) &  0.017 (0.017)  & 0.00 (0.06) & 33.673 (0.082) & $B,V$               & 2 \\
2007on  &  NGC 1404                       & 0.0062 & 0.588 (0.003) &  0.116 (0.012)  & 0.01 (0.01) & 31.453 (0.084) & $u,B,g,V,r,i,Y,J,H$ & 1 \\
2008R   &  NGC 1200                       & 0.0129 & 0.633 (0.006) &  0.120 (0.011)  & 0.08 (0.02) & 33.660 (0.080) & $u,B,g,V,r,i,Y,J,H$ & 1 \\
2008hs  &  NGC 0910                       & 0.0166 & 0.611 (0.006) &  0.075 (0.024)  & 0.04 (0.06) & 34.459 (0.093) & $B,V,r,i,J,H$       & 2,3,4,5 \\
2010Y   &  NGC 3392                       & 0.0113 & 0.658 (0.006) &  0.022 (0.016)  & 0.00 (0.06) & 33.861 (0.088) & $u,B,V,r,i,J,H$     & 2,3,4 \\
2011iv  &  NGC 1404                       & 0.0060 & 0.699 (0.007) &  0.031 (0.011)  & 0.07 (0.01) & 31.453 (0.084) & $u,B,g,V,r,i,Y,J,H$ & 1 \\
iPTF13ebh &  NGC 0890                     & 0.0125 & 0.636 (0.004) &  0.123 (0.009)  & 0.08 (0.02) & 33.296 (0.081) & $u,B,g,V,r,i,Y,J,H$ & 1 \\
2014bv  &  NGC 4386                       & 0.0057 & 0.621 (0.015) &  0.214 (0.022)  & 0.06 (0.03) & 32.427 (0.080) & $B,V$               & 2 \\
2015bp\tablenotemark{d}  &  NGC 5839      & 0.0047 & 0.703 (0.006) &  0.082 (0.013)  & 0.07 (0.01) & 32.369 (0.078) & $u,B,g,V,r,i,Y,J,H$ & 1 \\
CSP15aae\tablenotemark{d}  &  NGC 5490    & 0.0170 & 0.505 (0.004) &  0.454 (0.011)  & 0.16 (0.02) & 34.267 (0.080) & $u,B,g,V,r,i,Y,J,H$ & 1 \\
2016ajf &  NGC 1278                       & 0.0198 & 0.488 (0.015) &  0.556 (0.075)  & 0.14 (0.04) & 34.202 (0.106) & $B,V,r,i$           & 2,6 \\
\enddata
\tablecomments{Errors are shown in parentheses.}
\tablenotetext{a}{Values are K-corrected and corrected for Milky Way reddening. No correction has been applied for host galaxy reddening.}
\tablenotetext{b}{Estimate of the host galaxy reddening, $E(B-V)_{host}$, derived via the intrinsic color analysis described in detail by \cite{burns18}}
\tablenotetext{c}{Photometry references}
\tablenotetext{d}{CSP15aae = CSS150214:140955+1703155 (2015bo); 2015bp = SNhunt281}
\tablerefs{
(1) This paper; 
(2) \url{https://github.com/pbrown801/SOUSA/tree/master/data};
(3) \citet{hicken12};
(4) \citet{friedman15};
(5) \citet{stahl19};
(6) \citet{foley18}
}
\end{deluxetable*}
\end{longrotatetable}

\clearpage

\restartappendixnumbering

\section{SN~2007on and SN~2011iv in NGC~1404}
\label{sec:appendixC}

The transistional SNe~Ia 2007on (\sbv\ $= 0.588 \pm 0.003$) and 2011iv (\sbv\ $= 0.699 \pm 0.007$) were discovered four years apart in the 
Fornax Cluster E1 galaxy NGC~1404.  Both SNe were observed at optical and near-infrared wavelengths by the CSP \citep{gall2018} and extensively
modeled by \citet{ashall18} and \citet{mazzali18}.  As mentioned in \S\ref{sec:conclusions}, SN~2011iv was observed to be $0.30 \pm 0.02$~mag 
brighter than SN~2007on in the $B$-band after correction for color and light curve shape, and $0.20 \pm 0.01$~mag brighter in the $H$-band.

Another Fornax Cluster member that has produced four SNe~Ia since 1980 is the giant lenticular galaxy NGC~1316 (Fornax~A) \citep{stritzinger10}.
Three of these (1980N, 1981D, and 2006dd) were normal SNe~Ia with \sbv\ values in the range of 0.80--0.95, while the fourth was the very 
fast-declining SN~2006mr with \sbv\ $= 0.34$.

\begin{figure}[b]
\epsscale{.58}
\plotone{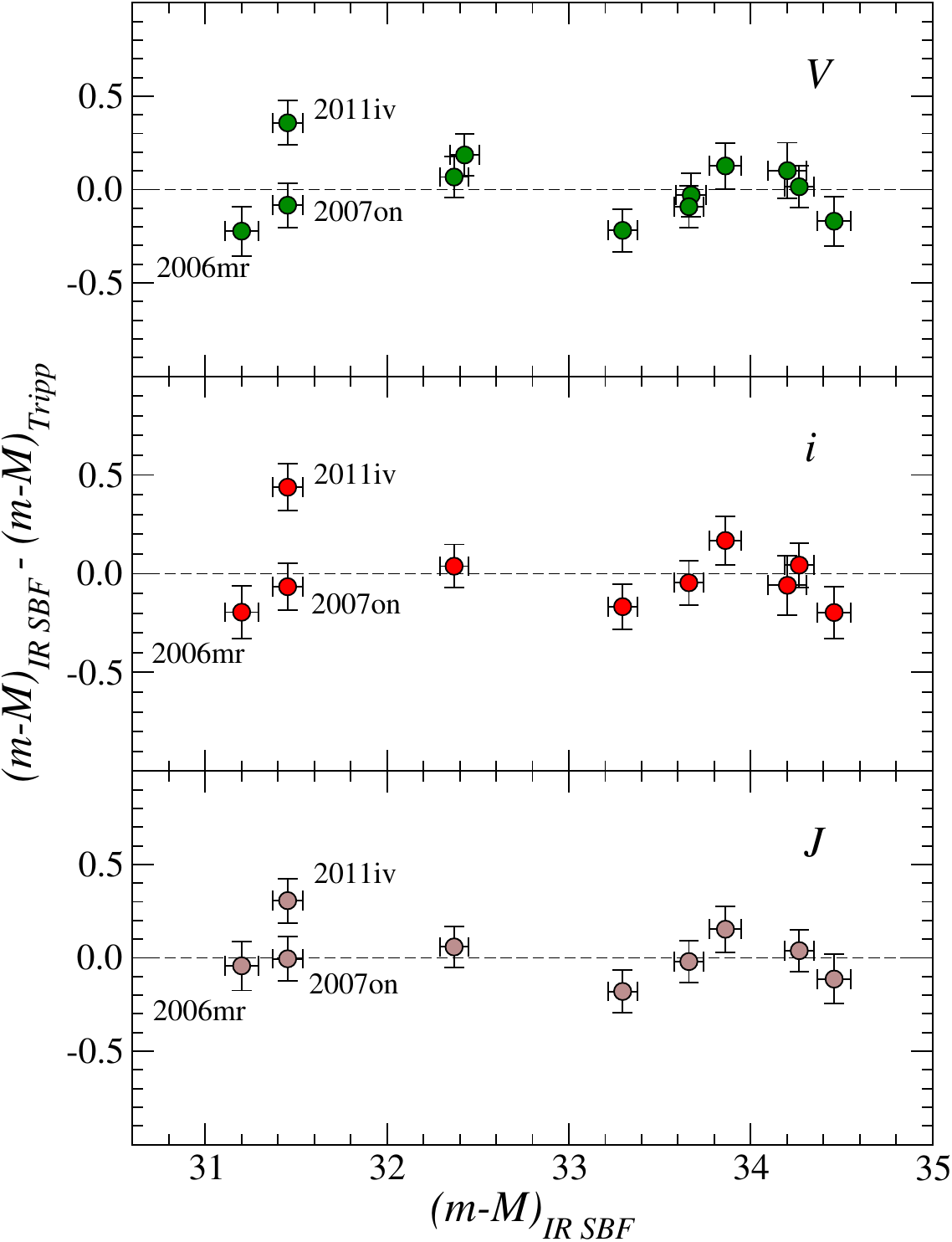}
\caption{Differences between the IR~SBF distance modulus, $(m-M)_{IR~SBF}$, and the observed distance modulus, $(m-M)_{Tripp}$, in the $V$, $i$, 
and $J$ bands using the quadratic ($N = 2$) Tripp method, plotted as a function of $(m-M)_{IR~SBF}$ for the calibrators.  The points corresponding to SN~2007on and 
SN~2011iv in NGC~1404 and SN~2006mr in NGC~1316 are labeled.}
\label{fig:delta_mu}
\end{figure}

Based on a comparison of TRGB and SBF distances to NGC~1316 and NGC~1404, and the SNe~Ia observed in both galaxies, \citet{hoyt21} argued
that SN~2011iv gives the more reliable distance to NGC~1404. However, when we compare the $H_0$ values calculated for these two SNe in our
analysis with those of the rest of the IR~SBF calibrators, SN~2011iv appears to be the more discrepant (see Figure~\ref{fig:delta_mu}).
This difference is traced to a 3.4$\sigma$ decrease in the IR~SBF distance moduli of NGC~1316 of $31.200 \pm 0.093$~mag as measured by 
\citet{garnavich23} with the IR channel of the Wide Field Camera~3 (WFC3/IR) compared to the \citet{blakeslee09} Advanced Camera for Surveys 
measurement of $31.583 \pm 0.065$~mag used by \citet{hoyt21}.

In principle, the three normal SNe~Ia that appeared in NGC~1316 can be used to provide an independent check of the distance of NGC~1316.
\citet{stritzinger10} employed the Tripp method and an assumed Hubble constant $H_0 = 72$~\kmsmpc\ to calculate a distance modulus of $31.248 
\pm 0.034~\rm{(statistical)} \pm 0.040~\rm{(systematic)}$~mag for these three SNe~Ia.  Adjusting this value to $H_0 = 75$~\kmsmpc\ gives 
$31.16$~mag.  Either of these values are consistent with the shorter WFC3/IR distance modulus for NGC~1316.

\clearpage

\restartappendixnumbering

\section{Priors and Corner Plots}
\label{sec:appendixD}

The priors employed for the MCMC solutions presented in \S\ref{sec:hubble} are as follows:
\begin{align*}
    -25<&M_B<14, \\
    -100<&p1<100, \\
    -100<&p2<100, \\
    -100<&\beta<100, \\
     0<&\sigma_{int}<100, \\
     0<&H_0 < 1000 
  \end{align*}

Corner plots for the linear ($N = 1$) solutions for the Tripp and Color methods are displayed in
Figures~\ref{fig:Tripp_corner_plot} and \ref{fig:Color_corner_plot}, respectively.

\begin{figure}[b]
\epsscale{1.1}
\plotone{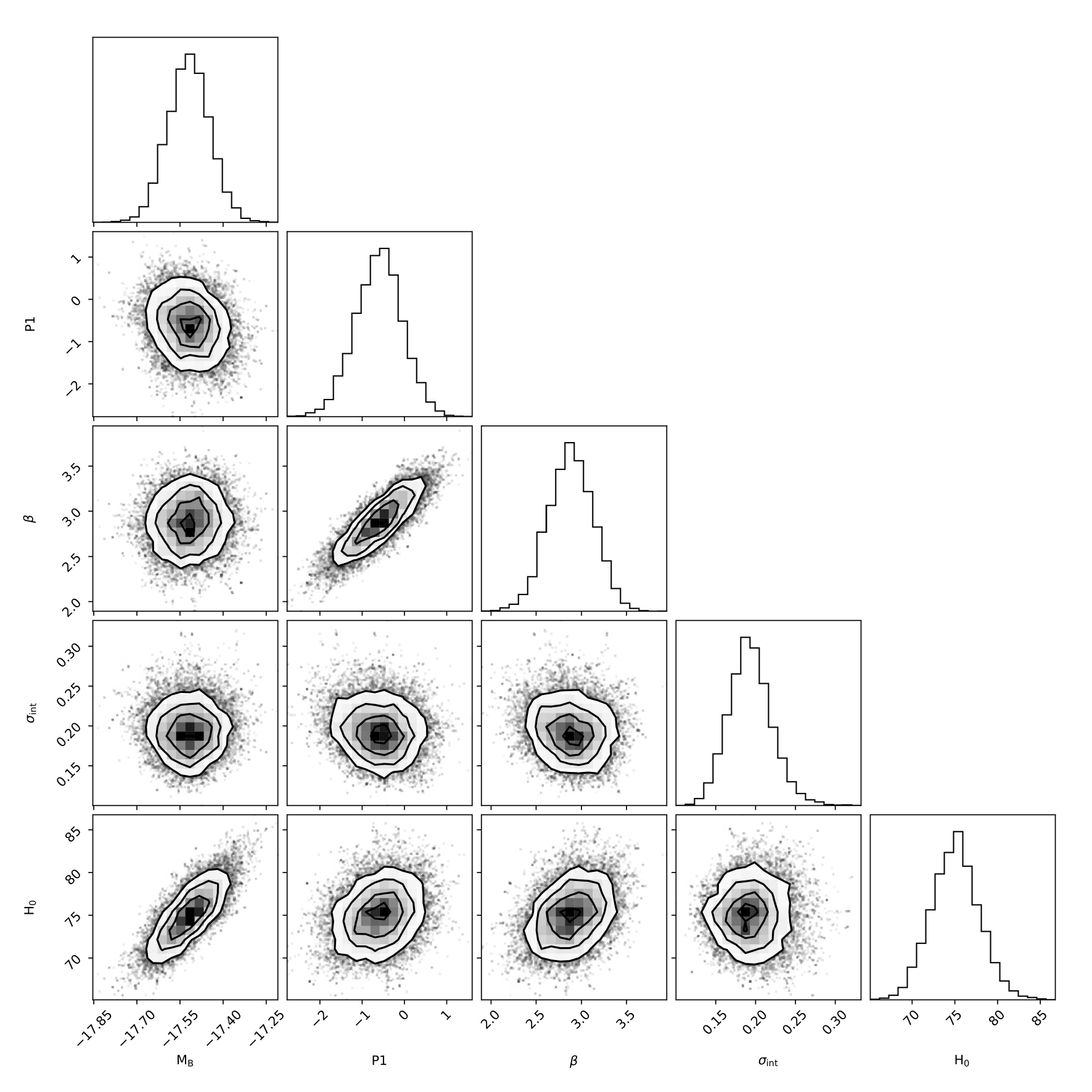}
\caption{Posterior distribution of MCMC fitting parameters in determining $H_0$ using the linear
($N = 1$) Tripp method solution for the B-band peak magnitudes.}
\label{fig:Tripp_corner_plot}
\end{figure}

\begin{figure}[b]
\epsscale{1.1}
\plotone{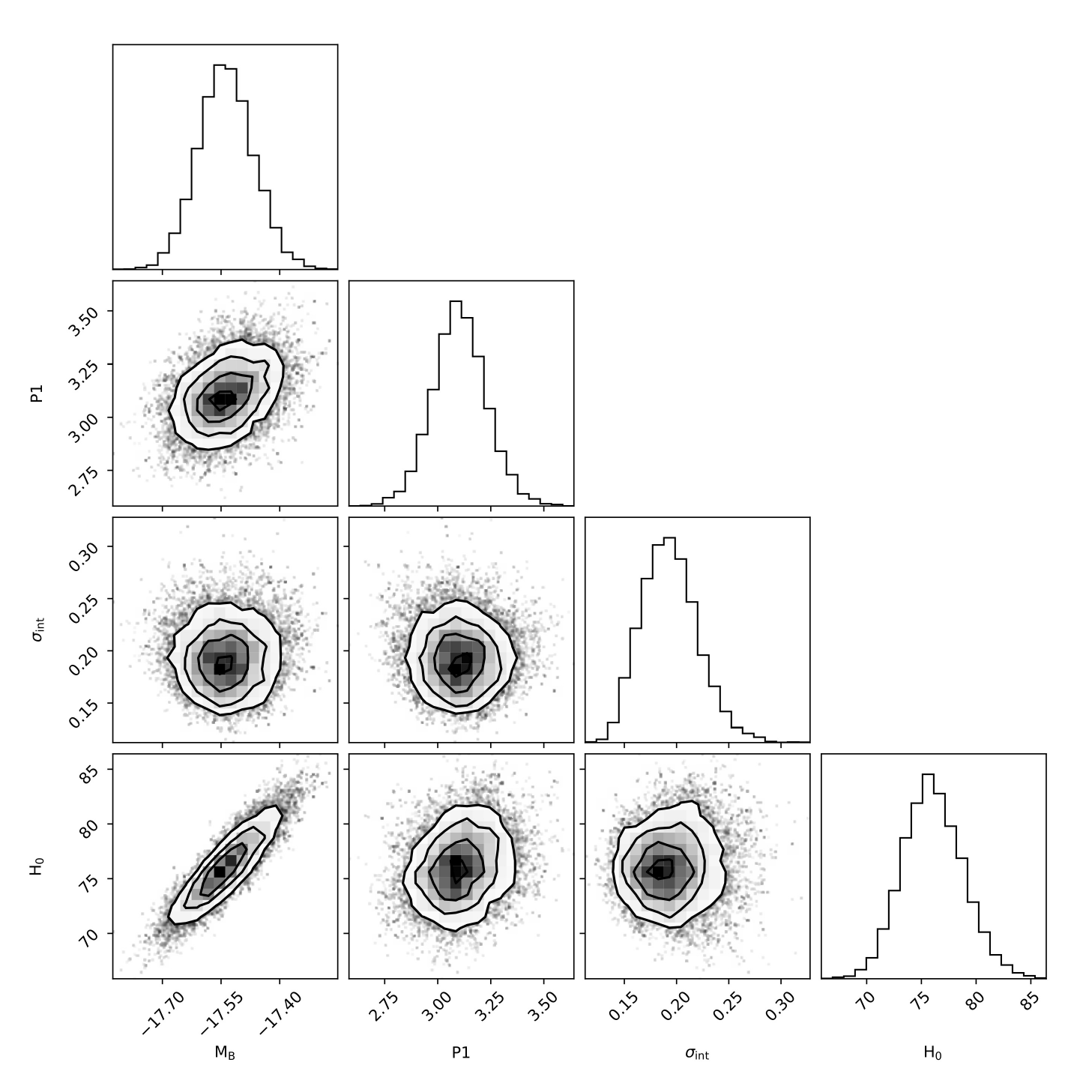}
\caption{Posterior distribution of MCMC fitting parameters in determining $H_0$ using the linear
($N = 1$) Color method solution for the B-band peak magnitudes.}
\label{fig:Color_corner_plot}
\end{figure}

\clearpage

\bibliography{ms_refs}{}
\bibliographystyle{aasjournal}

\end{document}